\documentclass[acmsmall,manuscript,screen]{acmart}

\AtBeginDocument{%
  \providecommand\BibTeX{{%
    \normalfont B\kern-0.5em{\scshape i\kern-0.25em b}\kern-0.8em\TeX}}}

\setcopyright{acmcopyright}
\copyrightyear{2021}
\acmYear{2021}
\acmDOI{10.1145/1122445.1122456}

\usepackage{amsmath}
\usepackage{todonotes}
\usepackage[inline]{enumitem}
\usepackage{tabularx}
\usepackage{multirow}
\usepackage{multicol}
\usepackage{booktabs}
\usepackage{subcaption}
\usepackage{float}
\newcolumntype{M}[1]{>{\centering\arraybackslash}m{#1}}

\begin{document}


\title{Conceptual Modeling and Artificial Intelligence: A Systematic Mapping Study}

\author{Dominik Bork}
\email{dominik.bork@tuwien.ac.at}
\orcid{0000-0001-8259-2297}
\affiliation{%
  \institution{TU Wien, Faculty of Informatics, Business Informatics Group}
  \streetaddress{Favoritenstr. 9-11}
  \postcode{1040}
  \city{Vienna}
  \country{Austria}
}

\author{Syed Juned Ali}
\email{syed.juned.ali@tuwien.ac.at}
\orcid{0000-0003-1221-0278}
\affiliation{%
  \institution{TU Wien, Faculty of Informatics, Business Informatics Group}
  \streetaddress{Favoritenstr. 9-11}
  \postcode{1040}
  \city{Vienna}
  \country{Austria}
  }

\author{Ben Roelens}
\email{ben.roelens@ou.nl}
\orcid{0000-0002-2443-8678}
\affiliation{%
  \institution{Open Universiteit, Faculty of Science, Information Science Group}
  \streetaddress{Valkenburgerweg 177}
  \postcode{6419 AT}
  \city{Heerlen}
  \country{The Netherlands}
  }
\affiliation{%
  \institution{Ghent University, Faculty of Economics and Business Administration, Business Informatics Research Group}
  \streetaddress{Tweekerkenstraat 2}
  \postcode{9000}
  \city{Ghent}
  \country{Belgium}
  }

\renewcommand{\shortauthors}{Bork et al.}

\begin{abstract}
In conceptual modeling (CM), humans apply abstraction to represent excerpts of reality for means of understanding and communication, and processing by machines. Artificial Intelligence (AI) is applied to vast amounts of data to automatically identify patterns or classify entities. While CM produces comprehensible and explicit knowledge representations, the outcome of AI algorithms often lacks these qualities while being able to extract knowledge from large and unstructured representations. Recently, a trend toward intertwining CM and AI emerged. This systematic mapping study shows how this interdisciplinary research field is structured, which mutual benefits are gained by the intertwining, and future research directions.
\end{abstract}



\maketitle

\section{Introduction}
\label{sec:intro}

In the last 60 years, Artificial Intelligence (AI) has become of strategic importance for organizations as it enables them to realize technological development in support of efficiently discovering and delivering innovative products and services \cite{AIWatch.20}. In particular, AI benefits from a large amount of available data to build intelligent applications and support fact-based decision making \cite{bork2020ER}. AI refers to \emph{software and/or hardware systems designed by humans that, given a complex goal, act in the physical or digital dimension by perceiving their environment through data acquisition, interpreting the collected structured or unstructured data, reasoning on the knowledge or processing the information derived from this data, and deciding the best action(s) to take to achieve the given goal} \cite[p.7]{HLEG.18}. AI systems are developed in domains like Machine Learning, Natural Language Processing (NLP), Computer Vision, Knowledge Representation, Automated Reasoning, Multi-agent Systems, etc.~\cite{GTIA.19}.
Although AI offers a vast range of opportunities for more efficient use of data in a digitized world, the quality of the data poses a significant challenge for organizations~\cite{Buxmann.21}. This implies that the data collection should be well embedded with the existing domain knowledge across different organization units to obtain consistent data sets. Furthermore, AI systems act as a black box that yields non-explainable results, which prevents a clear understanding and control of the followed process by human stakeholders. 

One can employ Conceptual Modeling (CM) techniques to tackle these challenges of AI systems as this domain has adopted a broad focus on enterprise-wide models and the complete business-to-IT domain~\cite{Winter.14}. CM enables the formal description of some aspects of the physical and social world around us for purposes of understanding and communication~\cite{Mylopoulos.92}. In the context of AI systems, four relevant CM purposes can be distinguished: $(i)$ representing a system~\cite{Mylopoulos.92,Bork.19,Olive.07,Thalheim.11}, $(ii)$ analyzing the properties of a system~\cite{Mylopoulos.92}, $(iii)$ (re-)designing a to-be system~\cite{Silva.15}, and $(iv)$ generating code to realize a system~\cite{Silva.15}. However, the value of CM is primarily appreciated in the collaboration between business architects and IT stakeholders, which results in a limited impact inside the organization~\cite{sandkuhl.18}. 

The combined use of CM and AI techniques defines the CMAI (i.e., Conceptual Modeling \& Artificial Intelligence) domain, which offers the opportunity to enhance the advantages and eliminate some of the disadvantages within the individual domains~\cite{Bork.21,Castellanos.21,Maass.21,Recker.21}. The combined value of CMAI is increased from two directions, the value added by CM to AI and, from the opposite direction, the value that AI adds to CM. This leads to the following characterization of research contributions in the CMAI domain~\cite{Bork.21}: $(i)$ \emph{Exaptation}, i.e., combining existing solutions from both fields to target a specific problem (e.g.,~\cite{Casini.12}); $(ii)$ \emph{CM-driven}, i.e., (re-)designing CM techniques combined with existing AI techniques to target a specific problem  (e.g.,~\cite{Bauer.01}); $(iii)$ \emph{AI-driven}, i.e., (re-)designing AI techniques combined with existing CM techniques to target a specific problem  (e.g.,~\cite{Boubekeur.20}); and $(iv)$ \emph{CM- \& AI-driven}, i.e., (re-)designing both AI and CM techniques to target a specific problem (e.g.,~\cite{bork2020ER}). 

Given the recency of the CMAI domain, getting a complete conceptualization is essential to increase the understanding of the domain and guide future research efforts. This conceptualization will be realized by performing a systematic mapping study (SMS) on the relevant literature~\cite{Wolny.20}. An SMS is preferred to a systematic literature review in this case as it enables to structure the overall body of knowledge through multiple categorization taxonomies~\cite{Wortmann.20}. Furthermore, these taxonomies support visual exploration to efficiently gain insights into the new CMAI domain. Literature reviews and mapping studies in the CMAI domain exist (see Sect.~\ref{sec:backgroundAndRelatedWork}). Still, existing works only focus on separated aspects of the AI and or CM domain, thereby preventing a comprehensive analysis of existing research and future research opportunities.

We automated the first part of the search process to enable such a comprehensive analysis of the CMAI domain. This included an automated extraction of the bibliometric data and the source files that resulted from the search queries' execution, which also enabled further filtering on the obtained paper sets regarding research discipline, publication type, and document length. Afterwards a manual analysis was feasible to assess the relevance of the content for the CMAI domain and to classify the papers according to the different categorization taxonomies, which we developed while being influenced from existing works: \emph{research type} (influenced by~\cite{Wortmann.20}), \emph{contribution type} (influenced by~\cite{Wortmann.20}), \emph{modeling purpose} (influenced by~\cite{Mylopoulos.92,Silva.15}), \emph{AI domain} (influenced by~\cite{Samoili.20}), and \emph{AI modeling value} (influenced by~\cite{Kinderen.15, AIWatch.20}). Sect.~\ref{sec:ResearchMethod} will provide more details on the development of the taxonomies. Afterward, we analyze how modeling languages and AI techniques have been combined throughout the last three decades. Furthermore, we present a web knowledge base\footnote{\url{http://me.big.tuwien.ac.at/cmai} currently only accessible to the reviewers. We will make it publicly available with the publication of this article.} that allows others to replicate the findings of this mapping study and to perform customized search queries in the CMAI domain. We will continuously update this web knowledge base to enable explorative search in the CMAI domain to the entire research community. Finally, implications for future CMAI research are derived based on these analyses.

This paper is structured as follows. Sect.~\ref{sec:backgroundAndRelatedWork} discusses related systematic literature reviews and mapping studies in the CM, AI, and CMAI domains. The research method that was followed for performing the mapping study is detailed in Sect.~\ref{sec:ResearchMethod}, while Sect.~\ref{sec:Findings} gives an answer to the proposed research questions. The CMAI web knowledge base is briefly introduced in Sect.~\ref{sec:WebSystemCMAI} and Sect.~\ref{sec:TrendsAndVision} presents the implications for future research. A discussion of the validity and reliability of the findings is presented in Sect.~\ref{sec:Validity}, after which this paper is concluded in Sect.~\ref{sec:Conclusion}.

\section{Related Work}
\label{sec:backgroundAndRelatedWork}
This section provides an overview of related work, i.e., related systematic literature reviews and mapping studies. As our research is positioned in the intersection of CM and AI, related works are first presented covering CM, followed by those works on AI. Eventually, we will report on existing works related to the combination of the two domains.

In the CM field, systematic literature/mapping studies exist that review the use of a particular (group of) modeling language(s). In work proposed by Wolny et al.~\cite{Wolny.20}, research on the Systems Modeling Language (SysML) between 2005 and 2017 is analyzed. This general-purpose architecture modeling language is an OMG standard for Systems Engineering applications~\cite{sysml}. In particular, Wolny et al.~\cite{Wolny.20} execute a mapping study of 579 publications to get an overview of existing research topics and groups, to identify publication trends, and to uncover missing links in the research domain. A broader view is adopted in other studies, which focus on the use of modeling languages for a particular purpose. Febrero et al.~\cite{Febrero.14} report on a mapping study of 503 papers to develop a taxonomy of modeling languages for software reliability. La Rosa et al.~\cite{laRosa.17} analyze 66 publications in a systematic literature review about business process variability modeling. The authors provide a comparative evaluation of the used languages to identify relevant modeling features, provide selection criteria among modeling approaches, and identify research gaps in the domain. In work by Silva et al.~\cite{Silva.15}, the use of CM for Model-Driven Engineering (MDE) is investigated, which includes the creation or automatic execution of software systems starting from these models. In particular, the authors conduct a literature survey to develop a unified model that identifies and relates the essential concepts of MDE. Wortmann et al.~\cite{Wortmann.20} present an analysis about the use of CM languages and techniques in a particular application domain, namely Industry 4.0. The extended mapping study reviews 408 publications to overview expected CM benefits, addressed Industry 4.0 concerns, CM languages used, applied research methods, involved countries and institutions, publication venues, and the temporal evolution of publication activities.

Research in the AI domain has been studied by Lu~\cite{Lu.19}, who performed a long-term mapping study of 7522 articles covering multiple angles in the AI domain: evolution through time, applications in industries, and challenges and future directions. Pournader et al.~\cite{Pournader.21} present a systematic analysis of AI applications in Supply Chain Management. In particular, 150 studies are analyzed to discuss the present and future state of the domain and the clusters of knowledge that currently exist.

Numerous systematic literature reviews and mapping studies are related to the CMAI domain by means of the fact that they cover at least a subset of the CMAI domain. The first category analyzes the interplay between CM and a selected AI technology or AI domain. For example, in~\cite{Raharjana.2021}, a systematic literature review is conducted to give an overview of 38 state-of-the-art research papers on implementing NLP in user stories. In particular, the authors focus on the contribution of NLP, the specific approaches that were used, and the challenges of using NLP in user story research. A similar focus is adopted by Zhao et al.~\cite{Zhao.21}, who analyze the use of NLP for Requirements Engineering. The authors present an SMS of about 404 relevant studies to conceptualize the studied combinations along with the following aspects: state of the literature, state of empirical research, research focus, tool development, and the usage of NLP technologies. Zaidi et al.~\cite{zaidi.21} review the use of CM for Machine Learning and vice versa. In total, ten papers are analyzed in detail to discuss the combined use of both and the challenges and opportunities related to this combination.

The second group of related works reviews the added value of applying one specific CM language in the AI domain. The work proposed by Goncalvez et al.~\cite{Goncalvez.18} discusses the use of iStar extensions, which is a goal modeling language for documenting requirements in software development. Although the literature review considers a wide range of application areas, 19 papers are identified covering the use of iStar to model agents, adaptive systems, autonomic systems, intelligent environments, and robotic systems. Liu et al.~\cite{Liu.2017} performed a literature review on the use of fuzzy Petri Nets to formalize knowledge representation and reasoning in rule-based expert systems. In total, 99 papers are analyzed to give an overview of existing theories and models, review the use of fuzzy Petri Nets in different application fields, and discuss current research trends and future research opportunities. 

The third category of related works reviews CMAI research in a particular application domain. In work proposed by Lee et al.~\cite{Lee.08}, the development and use of multi-agent modeling techniques and simulations are analyzed in the context of production design and development, production planning and control, and supply chain management. To this end, the authors review 114 applications to describe the state-of-the-art literature and to identify significant issues and a future vision for multi-agent modeling research.

The preceding discussion of related works shows the impressive amounts of work that analyze individual aspects of either CM or AI, or a very focused (e.g., application domain specific) combination of the two fields. To the best of our knowledge, the study we propose in this paper is the first of its kind by aiming for a broad, comprehensive, and inclusive scope on CM and AI that is not limited to particular modeling languages, AI techniques, or application domains.

\section{Research Method}
\label{sec:ResearchMethod}
\begin{figure}[t]
    \centering
    \includegraphics[width=.95\linewidth]{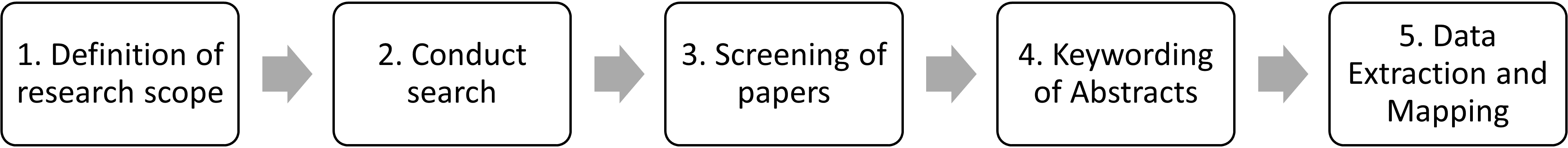}
    \caption{Phases of a SMS according to~\cite{Petersen.08}}
    \label{fig:research-steps}
\end{figure}

In the paper at hand, we report the findings of an SMS. Such studies aim \emph{"at exploring broad research areas by classifying the most representative studies in a particular subject and investigating generic research questions."}~\cite[p.\ 48:6]{Goncales.19}. Choosing an SMS over a Systematic Literature Review~\cite{Kitchenham.04} was motivated by the goal to classify and thematically analyze the vast body of literature on CM and AI instead of, e.g., identifying best practices~\cite{Kitchenham.11}. Furthermore, our search focus is broad, aimed to respond to generic research questions~\cite{Kitchenham.07,Kitchenham.11}. We followed the SMS research protocol proposed by Petersen et al.~\cite{Petersen.08} which is widely adopted and well recognized in the community (for exemplary applications, see~\cite{Zhao.21,Goncales.19,Meidan.18,Wortmann.20}). The protocol is composed of five phases (see Fig.~\ref{fig:research-steps}), which will be discussed in the remainder of this paragraph. 

\subsection{Phase 1: Definition of research scope} 
Typically for SMS, the research scope is broad, and the research interest, which is formalized by the research questions, is \emph{generic}~\cite{Kitchenham.11}. In our SMS, we thus concentrated on responding to the following research questions:

\begin{description}
\item[\emph{RQ-1: What is the state of the literature on CMAI?}] We want to know, who are the main contributors in the field, what types of reseaerch are contributed, and where is CMAI research published (i.e., the publication venues).
\item[\emph{RQ-2: Which type of research is performed in CMAI?}] We want to know which type of research is primarily reported on in CMAI (e.g., solution, evaluation, etc.) and also which contribution type the research offers to the scientific community (e.g., concepts, algorithms, etc.).
\item[\emph{RQ-3: Which modeling languages are combined with AI?}] We want to investigate which languages are used in combination with AI techniques.
\item[\emph{RQ-4: Which AI techniques are combined with CM?}] We want to explore which AI techniques are used in combination with CM languages.
\item[\emph{RQ-5: What are the mutual benefits of combining CM and AI?}] 
We want to know the benefits CM brings for AI and, vice versa, the benefits AI brings for CM.
\item[\emph{RQ-6: Which collaboration communities exist for CMAI?}] We want to know, what are the most active collaboration communities conducting CMAI research and on which topics they work on.
\end{description}

For many of the proposed research questions, we are interested in the current state and the longitudinal development (i.e., the trend analysis) of the CMAI research. 

\subsection{Phase 2: Conduct search} 
Following the guidelines proposed by Kitchenham et al.~\cite{Kitchenham.11}, an SMS search query should be \textit{generic} and guided by the topic of the study. Consequently, our search query combines the terms that best characterize \emph{Conceptual Modeling} and \emph{Artificial Intelligence}. The potential risk of having too many hits as a result of the broad search query can, to some extent, be mitigated by the fact that we are only interested in works intersecting both domains. Our study is based on the following logically structured search query that was also influenced by related previous SMSs (see Sect.~\ref{sec:backgroundAndRelatedWork}):

\begin{equation*} \label{eq1}
Q = (\lor CM_{i}) \land (\lor AI_{j}), where
\end{equation*}

\begin{center}
$CM_{i} \in$ \textit{$\{$"conceptual modeling" OR "metamodel" OR "meta-model" OR "domain specific language" OR} \newline
\textit{"modeling formalism" OR "modeling tool" OR "modeling language" OR "modeling method" OR} \newline
\textit{"model driven" OR "model-driven" OR "mde"$\}$
, and}\\
\end{center}

\begin{center}
$AI_{j} \in$ \textit{$\{$artificial intelligence" OR "ai" OR "machine learning" OR "ml" OR "deep learning" OR "dl"} \newline
\textit{"neural network" OR "genetic algorithm" OR "smart" OR "intelligent$\}$}.
\end{center}

In the executed query, we added the plural forms, e.g., "modeling languages" and the British English version, e.g., "modelling method" where applicable. We executed the query on the five well respected scientific databases \emph{ACM Digital Library}, \emph{IEEE Explore}, \emph{Science Direct}, \emph{Scopus}, and \emph{Web of Science}. We took the deliberate choice to rather be more generic with our search terms, e.g., including "smart" and "intelligent" for the AI search query part, instead of including not scientifically curated databases like Google Scholar.

\begin{figure}[t]
    \vspace{-.2cm}
    \centering
    \includegraphics[width=.78\linewidth]{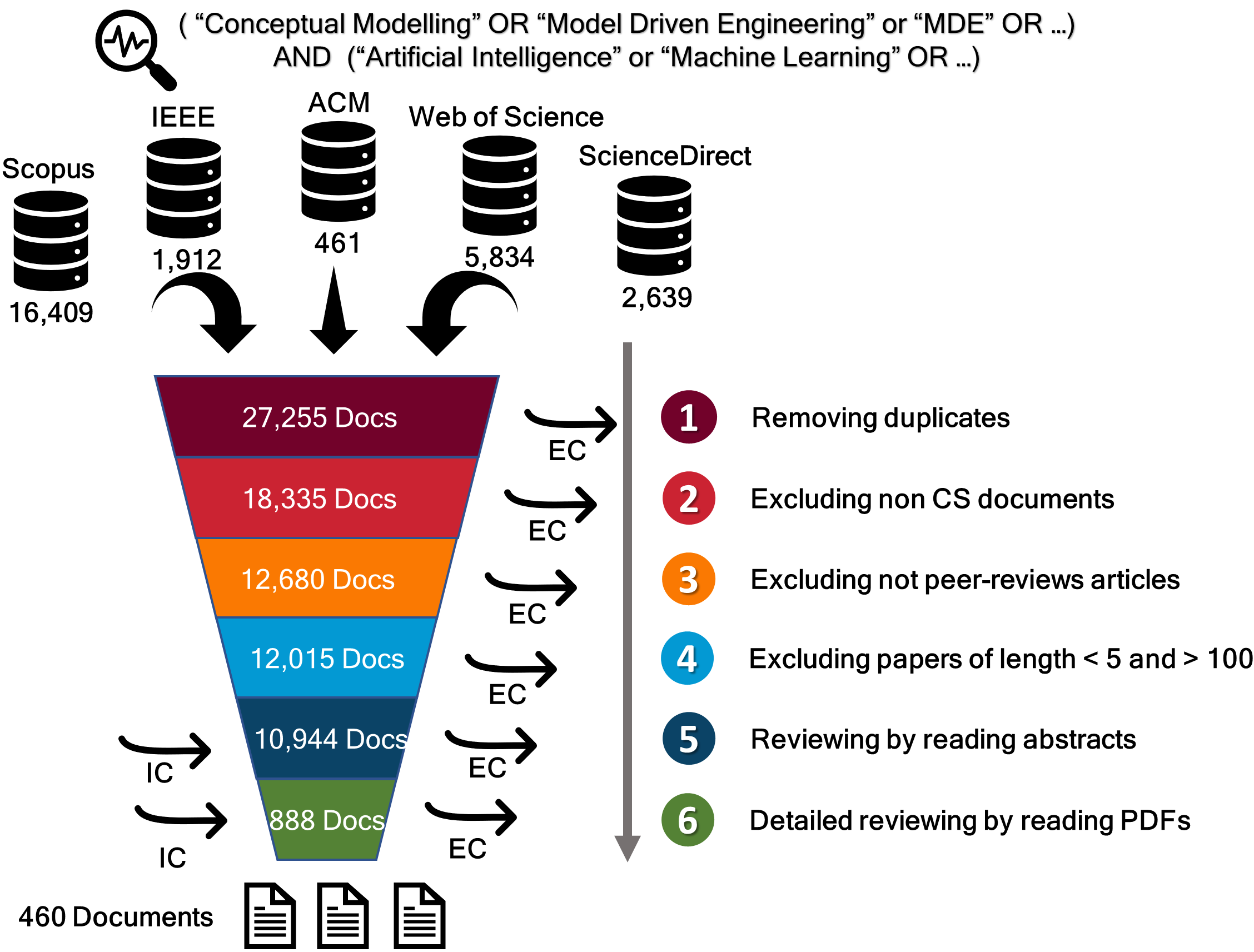}
    \caption{Document collection and major filtering steps.}
    \label{fig:filtering-steps}
\end{figure}

\subsection{Phase 3: Screening of papers} 
In the third step of an SMS, inclusion and exclusion criteria need to be defined to separate the relevant and irrelevant papers. For the study at hand, two following generic inclusion criteria were defined:
\begin{enumerate*}[label=\textbf{IC-\arabic{*}:}]
    \item Publications reporting on the combination of CM and AI (decided based on title, abstract, and keywords); 
    \item Peer-reviewed publications (i.e., articles in journals, conferences or workshops, and book chapters).
\end{enumerate*}

We applied the following list of exclusion criteria to identify and exclude irrelevant papers:
\begin{enumerate*}[label=\textbf{EC-\arabic{*}:}]
    \item Publications considering only CM or AI; 
    \item Publications not written in English; 
    \item Publications for which the full text is not available; 
    \item Extensions of already considered publications (we used the extended version); 
    \item Non-Peer-reviewed and non-scientific publications (e.g., editorials, tutorials, books, extended abstracts); 
    \item Publications not related to Computer Science (e.g., social sciences, chemistry); and 
    \item Publications with less than five or more than 100 pages.
\end{enumerate*}

The search was performed on 20th May 2021 on all five databases and yielded a total number of $27,255$ potentially relevant documents (see Fig.~\ref{fig:filtering-steps}). We retrieved the results from Scopus using the available API. The API response provided us with all the relevant information about the publications. We downloaded all the values from the online search portal for IEEE, ACM, Science Direct, and Web of Science. Once we had the downloaded citations, we transformed the BibTeX files to CSV. Once we had the CSVs of all the individual datasets, we removed the duplicates, which yielded $18,335$ remaining documents. Fig.~\ref{fig:filtering-steps} shows the number of results retrieved from the respective sources and the remaining number of unique hits after removal of the duplicates based on title and DOI (if present). We then removed all the publications that did not belong to the Computer Science discipline using the classification of the search databases (e.g., Science Direct). That resulted in $12,680$ remaining documents. Afterward, only conference papers, book chapters, and journal articles (IC-2) were considered. We thus filtered based on document type (IC-2 and EC-5), which yielded $12,680$ remaining documents. In the next step, we further applied a page length filter (EC-7), decreasing the number of remaining documents to $10,944$, thereby concluded the filtering based on bibliometric data and metadata.

In the final stage of the search process, we manually reviewed the remaining $10,944$ documents. We first downloaded the abstracts of all documents, read them, and decided on the relevance to our study. In this stage, $888$ documents remained relevant (IC-1, EC-1). We then tried to retrieve all full-texts (EC-3) of these $888$ documents, read them entirely, applied the remaining exclusion criteria (EC-2 and EC-4), and mapped them to our multi-faceted classification scheme. Eventually, $460$ documents were considered relevant and were mapped.

The generic and broad search query can explain the considerable effect of the filtering, which removed $98.32\%$ of the documents. For example, many documents were found that only considered CM in combination with the term 'smart,' i.e., modeling in smart cities~\cite{Bork.16SmartCity}. We also learned and consequently excluded studies that use the term 'meta-model' not in the sense we consider it relevant in the Computer Science domain. Many studies, e.g., from the mechanical engineering domain, use the term to relate to an artificial object that imitates the behavior of a physical thing for means of analysis and simulation~\cite{Fathi.14}.

\subsection{Phase 4: Keywording of Abstracts} 
\label{sec:ResearchMethod:Phase4}
In the fourth step of the SMS procedure proposed by Petersen et al.~\cite{Petersen.08}, the authors of the SMS read the abstracts of the relevant papers and \emph{"look for keywords and concepts that reflect the contribution of the paper"}~\cite[p.\ 4]{Petersen.08}. Once all authors of this paper read a representative sample of paper abstracts, we conducted a brainstorming session to identify relevant facets to characterize the CMAI research field and respond to the research questions introduced at the outset adequately. In the following, we will introduce the classification schemes, which will be used in Sect.~\ref{sec:Findings} to map the relevant literature into a coherent structure.

\subsubsection{Research type taxonomy}
\label{sec:ResearchType-taxonomy}
Papers can be categorized according to four different types of research: \textbf{Vision}, \textbf{Solution}, \textbf{Evaluation}, and \textbf{Experience}. This taxonomy is adopted from~\cite{Wortmann.20} and further refined based on~\cite{Petersen.08,Peffers.07} (see Table~\ref{tab:ResearchType-taxonomy}).

\begin{table}[h!]
\centering
\caption{Research Type Taxonomy}
\label{tab:ResearchType-taxonomy}
\footnotesize
\begin{tabularx}{\linewidth}{lX}
\toprule
\textbf{Research Type} & \textbf{Definition} \\
\midrule
Vision & Non-disruptive research agenda setting papers~\cite{Wortmann.20}.\\
\midrule
Solution & A solution for a problem is proposed; the solution can be either novel or a significant extension of an existing one. The potential benefits and the applicability of the solution are demonstrated by a small example or a solid line of argumentation~\cite{Petersen.08, Peffers.07}.\\
\midrule
Evaluation & Papers observing how a technique is implemented to solve the research problem (solution implementation) and measure the consequences of the implementation in terms of benefits and drawbacks (implementation evaluation)~\cite{Petersen.08, Peffers.07}.\\
\midrule
Experience & Explain what and how something has been done in practice, referring to the personal experience of the author(s)~\cite{Petersen.08}.\\
\bottomrule
\end{tabularx}
\vspace{-.1cm}
\end{table}

\subsubsection{Contribution type taxonomy}
\label{sec:ContributionType-taxonomy}
The literature will be classified according to five different types of research contributions: \textbf{Discussion}, \textbf{Concepts}, \textbf{Methods}, \textbf{Algorithms}, and \textbf{Tools}. The precise definition of this classification structure is given in Table~\ref{tab:ContributionType-taxonomy}, which is an extension based on~\cite{Petersen.08, Peffers.12, Wortmann.20}.

\begin{table}[h!]
\centering
\caption{Contribution Type Taxonomy}
\label{tab:ContributionType-taxonomy}
\footnotesize
\begin{tabularx}{\linewidth}{lX}
\toprule
\textbf{Contribution Type} & \textbf{Definition} \\
\midrule
Discussions & Papers contributing investigations without constructive contributions, e.g., reviews, commentaries, and opinions~\cite{Wortmann.20}.\\
\midrule
Concepts & Papers suggesting ways of thinking things, such as new metamodels, frameworks, or taxonomies that have been constructed from a set of statements, assertions, or other concepts~\cite{Peffers.12,Wortmann.20}.\\
\midrule
Methods & Papers suggesting new ways of doing things (e.g., applying existing models) by means of actionable instructions that are conceptual (not algorithmic)~\cite{Peffers.12,Wortmann.20}.\\
\midrule
Algorithms & Papers suggesting new automatic ways of computing (e.g., model transformation) or measuring things (i.e., metrics) by means of formal logical instructions~\cite{Peffers.12}.\\
\midrule
Tools & Papers presenting novel software tools, e.g., modeling tools~\cite{Petersen.08,Wortmann.20}.\\
\bottomrule
\end{tabularx}
\vspace{-.1cm}
\end{table}

\subsubsection{Modeling taxonomy}
\label{sec:Modeling-taxonomy}
The modeling taxonomy reflects the wide range of CM purposes included in our SMS. The taxonomy incorporates studies spanning from pure \textbf{Representation} purposes as stressed by, e.g., Mylopoulos~\cite{Mylopoulos.92}, Nelsen et al.~\cite{Nelson.12} amongst others~\cite{Bork.19,Olive.07,Thalheim.11}, to means of using conceptual models (i.e., the resulting artefact of the CM process~\cite{Mylopoulos.92}) to \textbf{Analyze} an existing system under study , e.g., by means of simulations of model queries~\cite{Smajevic.21}. Furthermore, the taxonomy includes the use of conceptual models to \textbf{(Re-)Design} or even automatically \textbf{Generate the Code} of a new system, i.e., treating models \emph{"as central artefacts in the software engineering process”}~\cite[p.\ 140]{Silva.15}.

\begin{table}[h!]
\centering
\caption{Modeling Purpose Taxonomy}
\label{tab:CM-taxonomy}
\footnotesize
\begin{tabularx}{\linewidth}{lX}
\toprule
\textbf{Modeling Purpose} & \textbf{Definition} \\
\midrule
Representation & Using a conceptual model to create an abstract representation of the system under study (i.e., descriptive modeling).\\
\midrule
Analysis & Using a conceptual model to analyze properties of the system under study employing, e.g., simulations or queries.\\
\midrule
(Re-)Design & Using a conceptual model to (re-)design a future to-be version of the system under study (i.e., prescriptive modeling).\\
\midrule
Code Generation & Using a conceptual model to generate (parts of) the code that can be executed to realize a (software) system.\\
\bottomrule
\end{tabularx}
\end{table}

\subsubsection{AI Modeling Value Taxonomy}
\label{sec:AIModelingValue-taxonomy}
In this taxonomy, we are interested in analyzing the benefit/value that AI brings for CM. The two main groups of stakeholders of conceptual models are the creators and the users of a model~\cite{Kinderen.15}. With regard to \textbf{model creation}, AI can help to \textbf{(semi-)automate} or even \textbf{automate} parts of the development process by data acquisition and interpretation~\cite{AIWatch.20}. On the other hand, AI may facilitate \textbf{model use} by \textbf{analyzing} model data and providing relevant information, and/or \textbf{processing} the model content to identify action(s) that achieve the goals of model users~\cite{AIWatch.20}.

\begin{table}[h!]
\centering
\caption{AI Modeling Value Taxonomy}
\label{tab:AImodelingvalue-taxonomy}
\footnotesize
\begin{tabularx}{\linewidth}{llX}
\toprule
\textbf{Value Domain} & \textbf{Value Subdomain} & \textbf{Definition} \\
\midrule
Model Creation & Semi-automated & A combination of the manual and automated creation of conceptual models, e.g., using AI for data acquisition and interpretation.\\
\cmidrule{2-3}
 & Automated & The automated creation of conceptual models, e.g., using AI for data acquisition and interpretation. \\
 \midrule
Model Use & Analyzing & Examining model data and providing relevant information by the use of AI.\\
\cmidrule{2-3}
 & Processing & Processing model data with the use of AI, e.g., to transform the model or to execute the model.\\
\bottomrule
\end{tabularx}
\end{table}

\subsubsection{AI Taxonomy}
\label{sec:AI-taxonomy}
The AI field is even more extensive than CM and composes many techniques applicable to different contexts. It is therefore not surprising that there exists not yet a commonly agreed-upon definition of AI. Existing definitions are furthermore very diverse spanning the metaphors thinking humanly, acting humanly, thinking rationally, and acting rationally (cf.~\cite{RussellNorvig.20,Samoili.20} for an overview of selected AI definitions). For our SMS, we adopted a recently proposed AI taxonomy that composes eight AI domains and 16 AI subdomains~\cite{Samoili.20} (see Table~\ref{tab:AI-taxonomy}). 
The taxonomy has been proposed in response to the lack of a common definition of AI and to account for the heterogeneity of AI research. It, therefore, is expressive enough to cover the broad scope we want to apply in our SMS.

\begin{table}
\centering
\caption{AI Taxonomy}
\label{tab:AI-taxonomy}
\footnotesize
\begin{tabularx}{\linewidth}{lp{2.9cm}X}
\toprule
\textbf{AI Domain} & \textbf{AI Subdomain} & \textbf{Definition} \\
\midrule
Reasoning & Knowledge Representation & Encoding human knowledge and reasoning into a symbolic language that enables it to be processed by information systems~\cite{Swain.13}. \\
\cmidrule{2-3}
 & Automated Reasoning & The ability to automatically make inferences on a computing system~\cite{Keet.13}.\\
 \cmidrule{2-3}
 & Common Sense Reasoning & Using the set of background information (facts) that an individual is expected to know in reasoning~\cite{Farhang.19}.\\
 \midrule
Planning & Planning and Scheduling & The realization of strategies or action sequences, typically for execution by intelligent agents, autonomous robots, and unmanned vehicles~\cite{AIWatch.20}. \\
\cmidrule{2-3}
 & Searching & Search is a universal problem-solving mechanism in AI.\\
 \cmidrule{2-3}
 & Optimization & Finding the optimal solution of a problem given certain constraints.\\
 \midrule
Learning & Machine Learning & Ability of systems to automatically learn, decide, predict, adapt, and react to changes, improving from experience, without being explicitly programmed~\cite{AIWatch.20}.\\
\midrule
Communication & Natural Language Processing & A machine’s ability to identify, process, understand and/or generate information in written and spoken human communications~\cite{AIWatch.20}.\\
\midrule
Perception & Computer Vision & An interdisciplinary field that uses Computer Science techniques to analyze and understand digital images and videos. Computer vision tasks include e.g., object recognition, event detection, motion detection, and object tracking~\cite{Farhang.19}.\\
\cmidrule{2-3}
 & Audio Processing & Deriving meaningful information from audio signals -- and taking actions or making recommendations based on that information. \\
 \midrule
 Integration and Interaction & Multi-agent systems & An environment composed of objects and agents (the agents being the only ones to act), relations between all the entities, a set of operations that can be performed by the entities, and the changes of the universe in time and due to these actions~\cite{Weiss.99}.\\
 \cmidrule{2-3}
 & Robotics and Automation & Activities related to assist or substitute human activity, or to enable actions that are not humanly possible (e.g., medical robots), to optimize technical limitations, labour or production costs~\cite{AIWatch.20}.\\
 \cmidrule{2-3}
 & Connected and Automated vehicles & Regards technologies of autonomous vehicles, connected vehicles and driver assistance systems, considering all automation levels and all communication technologies~\cite{AIWatch.20}.\\
 \midrule
Services & AI Services & Refers to any infrastructure, software and platform (e.g., cognitive computing, ML frameworks, bots and virtual assistants, etc.) provided as (serverless) services or applications, possibly in the cloud, which are available off the shelf and executed on demand, reducing the management of complex infrastructures~\cite{AIWatch.20}.\\
\midrule
Ethics and Philosophy & AI Ethics & Compliance with ethical principles and values, including applicable regulation~\cite{AIWatch.20}.\\
\cmidrule{2-3}
 & Philosophy of AI & Implications of AI for knowledge and understanding of intelligence, ethics, consciousness, epistemology, and free will~\cite{AIWatch.20}.\\
\bottomrule
\end{tabularx}
\end{table}

\subsection{Phase 5: Data Extraction and Mapping}
The final phase of our SMS, as proposed by Petersen et al.~\cite{Petersen.08}, is the data extraction and mapping phase. We set up a shared Google Docs spreadsheet to collaborate remotely. The taxonomies were pre-configured in the spreadsheet to ease the mapping process. We conducted weekly meetings to discuss the progress, streamline the taxonomies, and develop a coherent and consistent evaluation of the papers. Once we finished the mapping, we applied several data processing steps to generate insights and visualizations. All findings of this step will be presented in detail in the Sect.~\ref{sec:Findings}.

\section{Findings}
\label{sec:Findings}

In this section, we present the findings of our SMS. The research questions will be responded to by explaining the results of mapping the remaining relevant papers to the individual taxonomies introduced at the outset. As our SMS comprises 460 eventually relevant papers, some results of categories like the used CM language or the countries the study authors are affiliated with exceed the space and hamper visual representation when presenting all individual results. In such cases, we will define a threshold that constrains the values shown in this paper. A comprehensive analysis of all results is provided in the accompanying web knowledge base (see Sect.~\ref{sec:WebSystemCMAI}). The presentation of the findings is structured as follows: Sect.~\ref{sec:findings:BibliometricAnalysis} provides a bibliometric analysis, reporting, e.g., on a longitudinal overview of CMAI research and the most active contributors to the field; Sect.~\ref{sec:findings:ContentAnalysis} then reports on the content of the CMAI contributions, e.g., how often a CM language is used in combination with AI techniques; and eventually, Sect.~\ref{sec:findings:Community} reports on an analysis of the CMAI research communities.

\subsection{Bibliometric Analysis}
\label{sec:findings:BibliometricAnalysis}
As a response to RQ-1, this section describes the current state of the literature on CMAI research, including an analysis of $(i)$ the number of papers and the $(ii)$ the publication types over time. Moreover, this also includes an analysis of the $(iii)$ main contributors to CMAI research and $(iv)$ the primary publication venues used.

\subsubsection{Longitudinal overview of CMAI research}
\label{sec:LongitudinalAnalysis}
\begin{figure}[h!]
    \centering
    \vspace{-.3cm}
    \includegraphics[width=.9\linewidth]{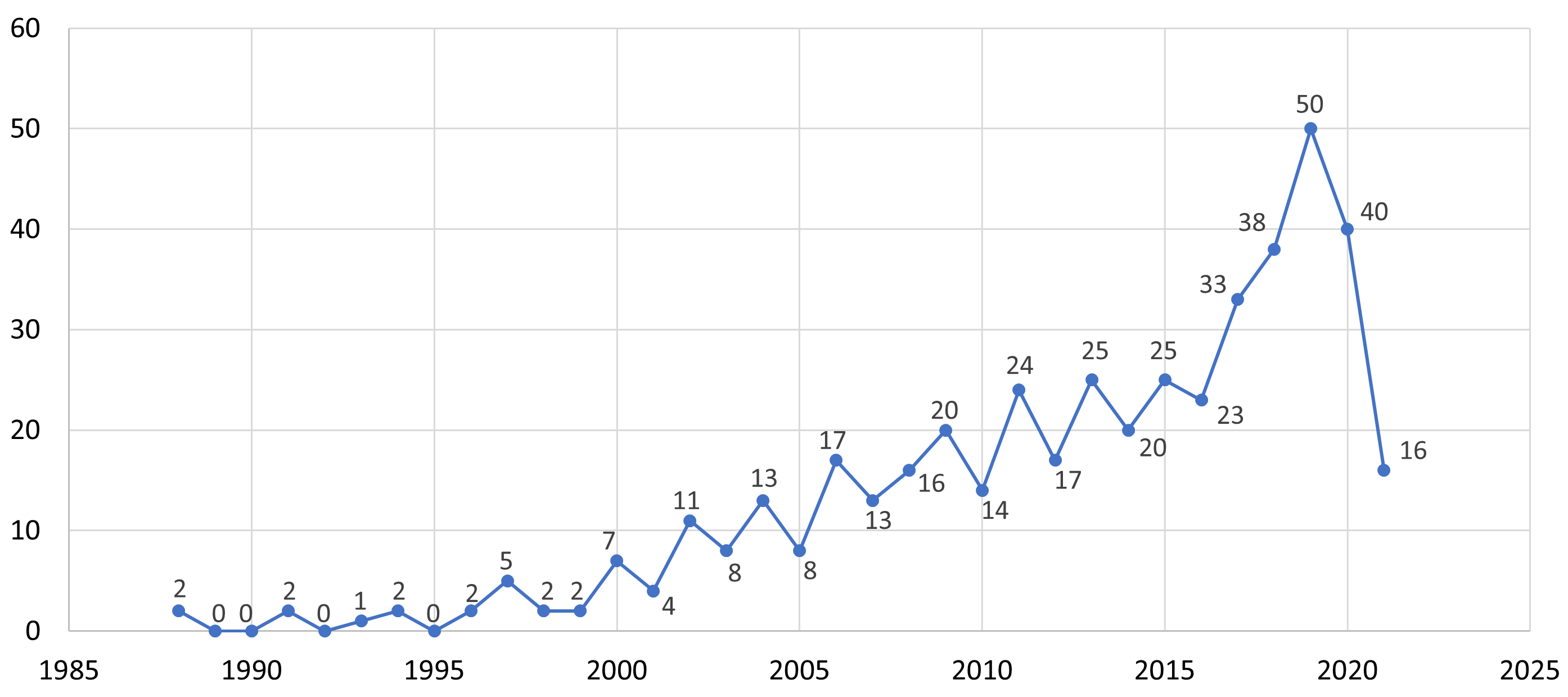}
    \caption{Number of publications per year until May 2021.}
    \label{fig:papers-per-year}
    \vspace{-.3cm}
\end{figure}

Fig.~\ref{fig:papers-per-year} depicts the number of papers per year, which shows a zigzag pattern with a positive trend. Note that the figure for 2021 only shows the results until the 20th of May. The first two papers in the CMAI domain date back to 1988, and the number increased to a maximum of 50 in 2019. The domain took off slowly as 15\% of the publications were published between 1988 and 2005. Afterward, the growth has steadily accelerated, which resulted in 48.9\% of the publications since 2015. Hence, it can be concluded that there has been a growing interest in the combined use of CM and AI techniques, which shows the increasing relevance of the CMAI domain in the last three decades.

\begin{figure}[H]
    \centering
    \includegraphics[width=.95\linewidth]{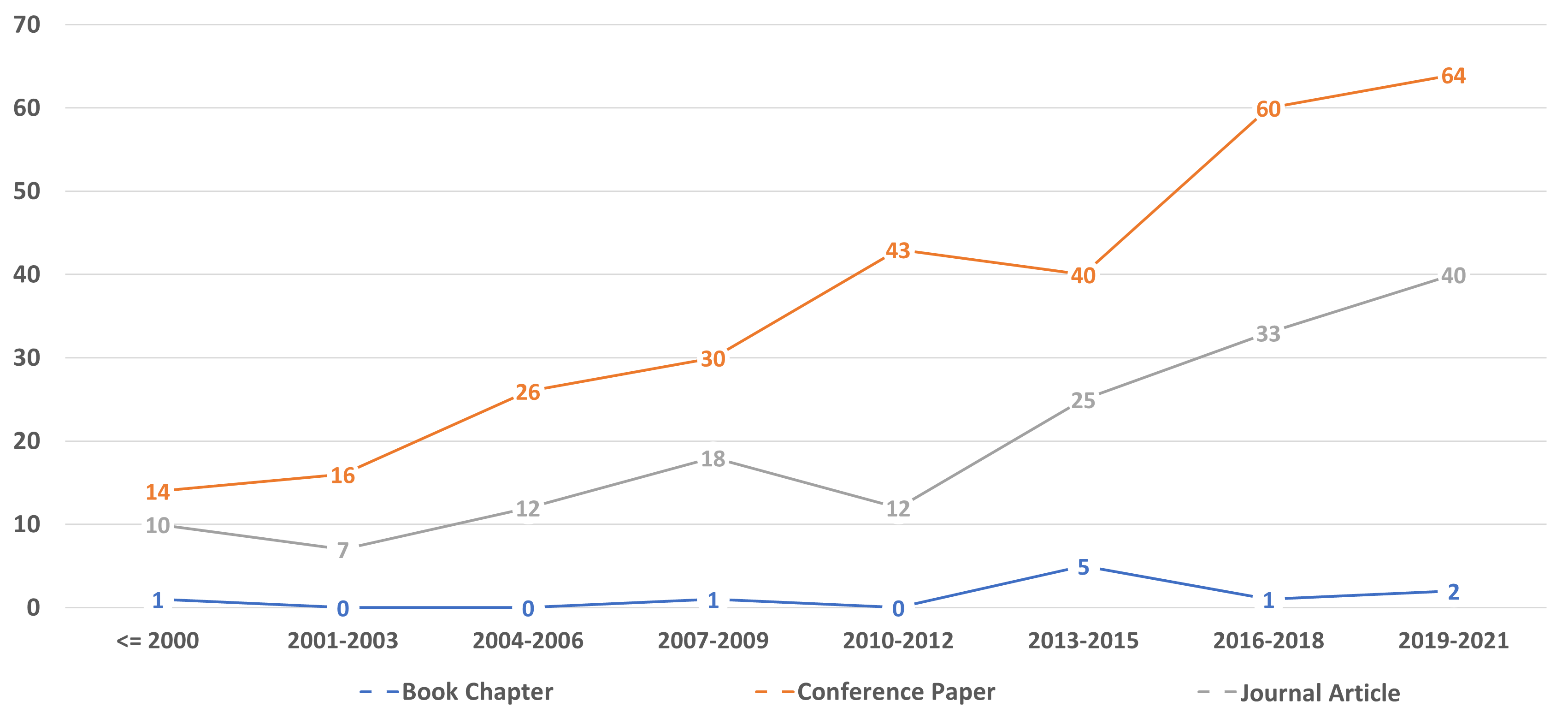}
    \caption{Number of publications classified based on publication type and grouped by intervals of three years.}
    \label{fig:Document-Type-Trend}
\end{figure}

Fig.~\ref{fig:Document-Type-Trend} shows that the work in the CMAI domain is currently mainly published as conference papers (i.e., 60.4\% of publications in the period 2019-2021) and journal articles (i.e., 37.7\% of publications in the period 2019-2021). This mutual ratio has remained relatively stable over time, with a mean of 63.7\% conference papers (i.e., 293 of 460) and 34.1\% (i.e., 157 of 460) journal articles since 1988. The absolute number of book chapters fluctuates between 0 and 5 over the years.

\subsubsection{Main contributors}
\label{sec:contributors}
\begin{figure}[ht!]
    \centering
    \vspace{-.3cm}
    \includegraphics[width=.95\linewidth]{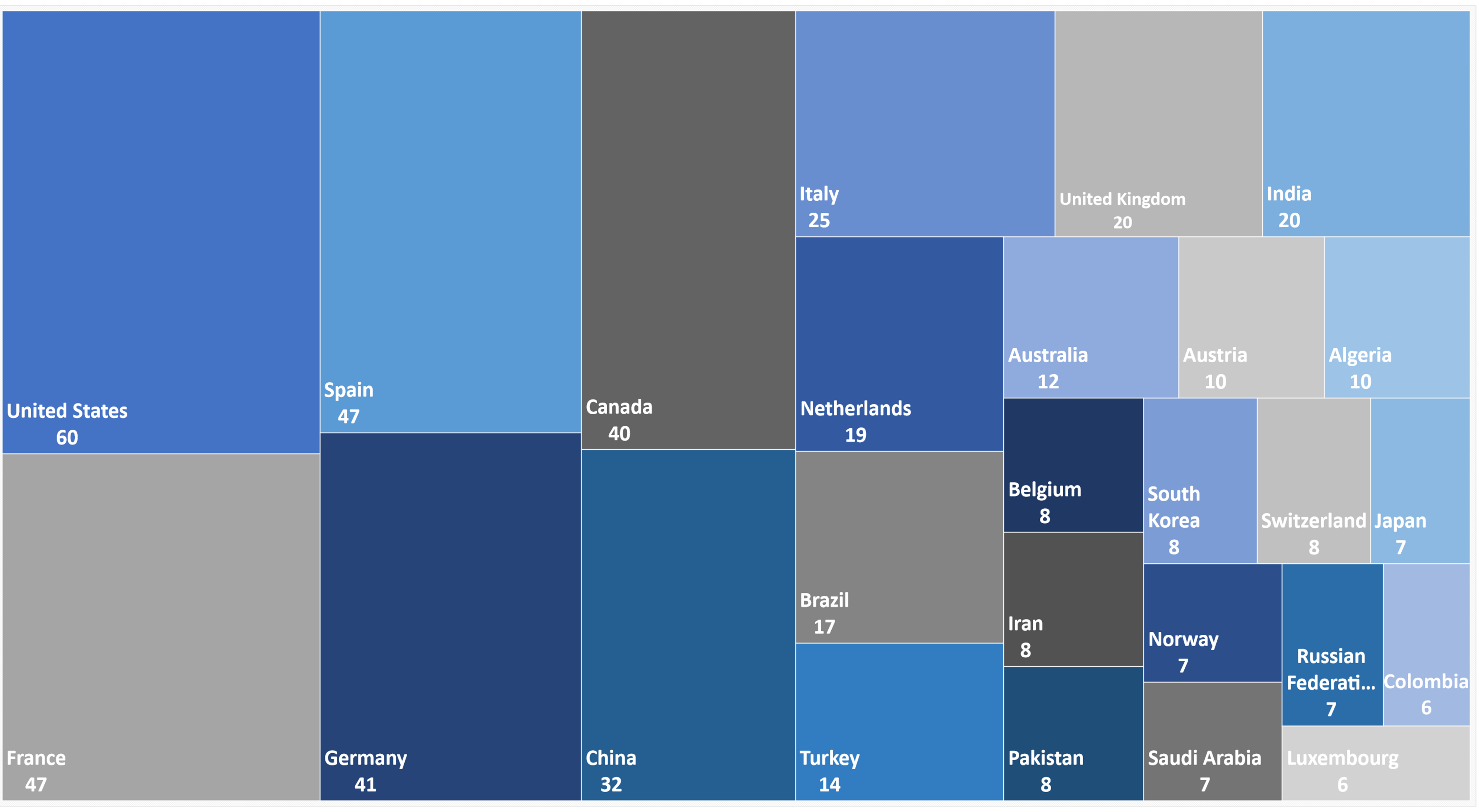}
    \caption{Publications share by countries with more than five publications (representing 88.69\% of the survey data).}
    \label{fig:papers-per-country}
    \vspace{-.3cm}
\end{figure}

The publication share per country is given in Fig.~\ref{fig:papers-per-country}, which includes all countries with more than five contributions. To identify the countries, we analyzed the geographical location of all authors' affiliations. This resulted in 26 countries for the $1,186$ authors in the dataset. The five most contributing countries, which account for 39.2\% of the publications, are distributed between North America and Europe, including the United States (60 publications), France (47 publications), Spain (47 publications), Germany (41 publications), and Canada (40 publications).

\begin{table}[h!]
\centering
\caption{Five most active institutions}
\label{tab:top-institutions}
\footnotesize
\begin{tabularx}{\linewidth}{Xp{.2\linewidth}p{.15\linewidth}}
\toprule
\textbf{Institution} &\textbf{Country} &\textbf{No. of Papers} \\
    \midrule
     Complutense University of Madrid & Spain & 14 \\
     \midrule
     University of Toronto & Canada & 11 \\
     \midrule
     Ege University & Turkey & 10 \\
     \midrule
     McGill University & Canada & 6 \\
     \midrule
     University of Montreal & Canada & 6 \\
    \bottomrule
\end{tabularx}
\end{table}

Table~\ref{tab:top-institutions} analyzes the institutions that are most active in CMAI research. The leading institution that published relevant research is the Complutense University of Madrid in Spain (14 papers), particularly the Research Group on Agent based Social and Interdisciplinary Applications (GRASIA) of the Department of Software Engineering and Artificial Intelligence. Notably, three Canadian universities are part of the five most active institutions (i.e., the University of Toronto, McGill University, and the University of Montreal), besides the Ege University of Turkey.

\subsubsection{Publication channels}
\label{sec:publicationchannels}
Table~\ref{tab:top-venues} gives an overview of the most popular journals and conferences that published CMAI research. These provide a good starting point for authors searching for channels to distribute CMAI research. Nevertheless, the journals and conferences in Table~\ref{tab:top-venues} only account for 16.5\% of the publications in the dataset, which illustrates the fragmentation that currently still exists in the domain. This is not surprising, as this SMS is the first effort to provide a broad overview of the CMAI field. Regarding the journals, Expert Systems with Applications is the most popular by disseminating ten papers since 2005. The conference that published most CMAI research (i.e., 12 publications) is the International Conference on Model Driven Engineering Languages and Systems (MoDELS). 

\begin{table}[h!]
\centering
\vspace{-.2cm}
\caption{The most popular journals and conferences}
\label{tab:top-venues}
\footnotesize
\begin{tabularx}{\linewidth}{Xp{.15\linewidth}}
\toprule
\textbf{Name} & \textbf{No. of Papers} \\
    \midrule
    \multicolumn{2}{c}{\textbf{Journals}} \\
    \midrule
     Expert Systems with Applications & 10 \\
     \midrule
     Journal of Systems and Software & 5 \\
     \midrule
     Information and Software Technology & 4 \\
     \midrule
     Automated Software Engineering & 3 \\
     \midrule
     Engineering Applications of Artificial Intelligence & 3 \\
     \midrule
     International Journal of Software Engineering and Knowledge Engineering & 3 \\
     \midrule
     Robotics and Computer-Integrated Manufacturing & 3 \\
     \midrule
     \multicolumn{2}{c}{\textbf{Conferences}} \\
     \midrule
     International Conference on Model Driven Engineering Languages and Systems, MODELS & 12 \\
     \midrule
     International Conference on Conceptual Modeling, ER & 10 \\
     \midrule
     International Conference on Model Driven Engineering Languages and Systems Companion, MODELS-C & 8 \\
     \midrule
     International Conference on Tools with Artificial Intelligence, ICTAI & 8 \\
     \midrule
     International Workshop on Agent-Oriented Software Engineering, AOSE & 7 \\
    \bottomrule
\end{tabularx}
\vspace{-.3cm}
\end{table}

\subsection{Content Analysis}
\label{sec:findings:ContentAnalysis}
As a response to RQ-2 through RQ-5, this section contributes the content analysis of the surveyed CMAI research. This includes $(i)$ a meta-analysis of the research and contribution types (Sect.~\ref{sec:ContentAnalysis:Meta}); $(ii)$ an analysis of the CM facet (modeling languages and modeling purpose) in Sect.~\ref{sec:ContentAnalysis:ModelingTaxonomy}; $(iii)$ an analysis of the AI facet in Sect.~\ref{sec:ContentAnalysis:AITaxonomy}; and $(iv)$ an analysis of the combined use of CM and AI in Sect.~\ref{sec:ContentAnalysis:CMAI}.

\subsubsection{Meta Analysis}
\label{sec:ContentAnalysis:Meta}
In a meta analysis, we are interested to classify the research type (see Table~\ref{tab:ResearchType-taxonomy}) and the contribution type (see Table~\ref{tab:ContributionType-taxonomy}) of CMAI research.

\begin{figure}[h!]
    \centering
    \vspace{-.1cm}
    \includegraphics[width=.8\linewidth]{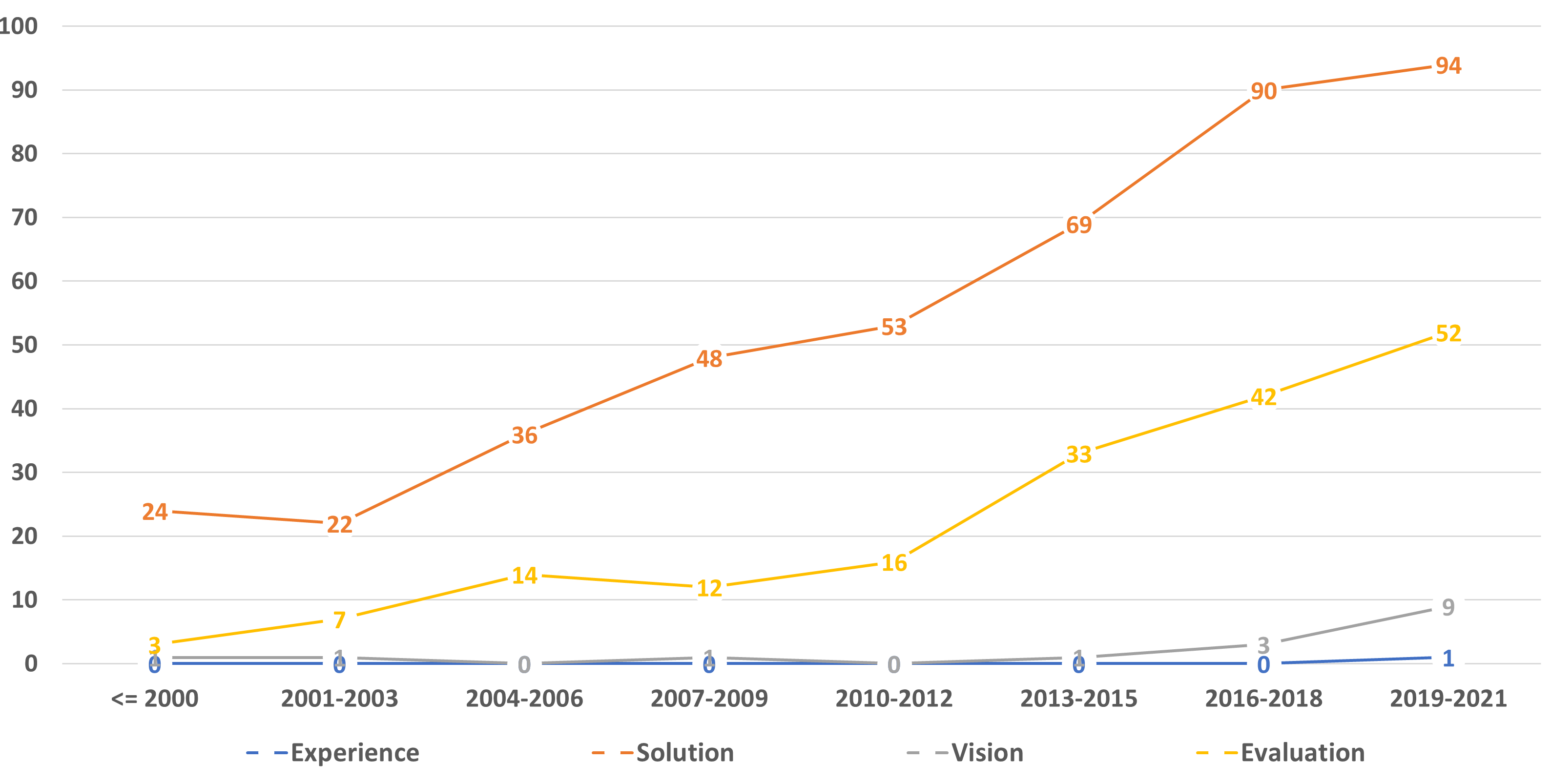}
    \caption{Number of publications classified based on research type and grouped by intervals of three years.}
    \label{fig:research-type-trend}
    \vspace{-.3cm}
\end{figure}

\paragraph{Research type.}
\label{sec:ContentAnalysis:Meta:ResearchType}
The majority of CMAI research (i.e., 436 occurrences) is oriented towards solving a particular problem and demonstrating the applicability of a proposed solution (see Fig.~\ref{fig:research-type-trend}). The second most occurring research type is evaluation, which was observed in 179 cases. In 172 papers (i.e., 37.4\% of the dataset), the proposed solution is supplemented by an evaluation, which is in line with the Design Science Research (DSR) methodology~\cite{gregor2013positioning, Peffers.07}. DSR research proposes a solution with an evaluation to complete the iterative build-and-evaluate loop, i.e., DSR research includes the development of a new artifact and the use of this artifact to solve a particular problem~\cite{Peffers.07}. Vision and experience papers remain a niche in the CMAI domain with respectively 16 and 1 occurrence(s) since 1988.

\begin{figure}[h!]
    \centering
    \includegraphics[width=.95\linewidth]{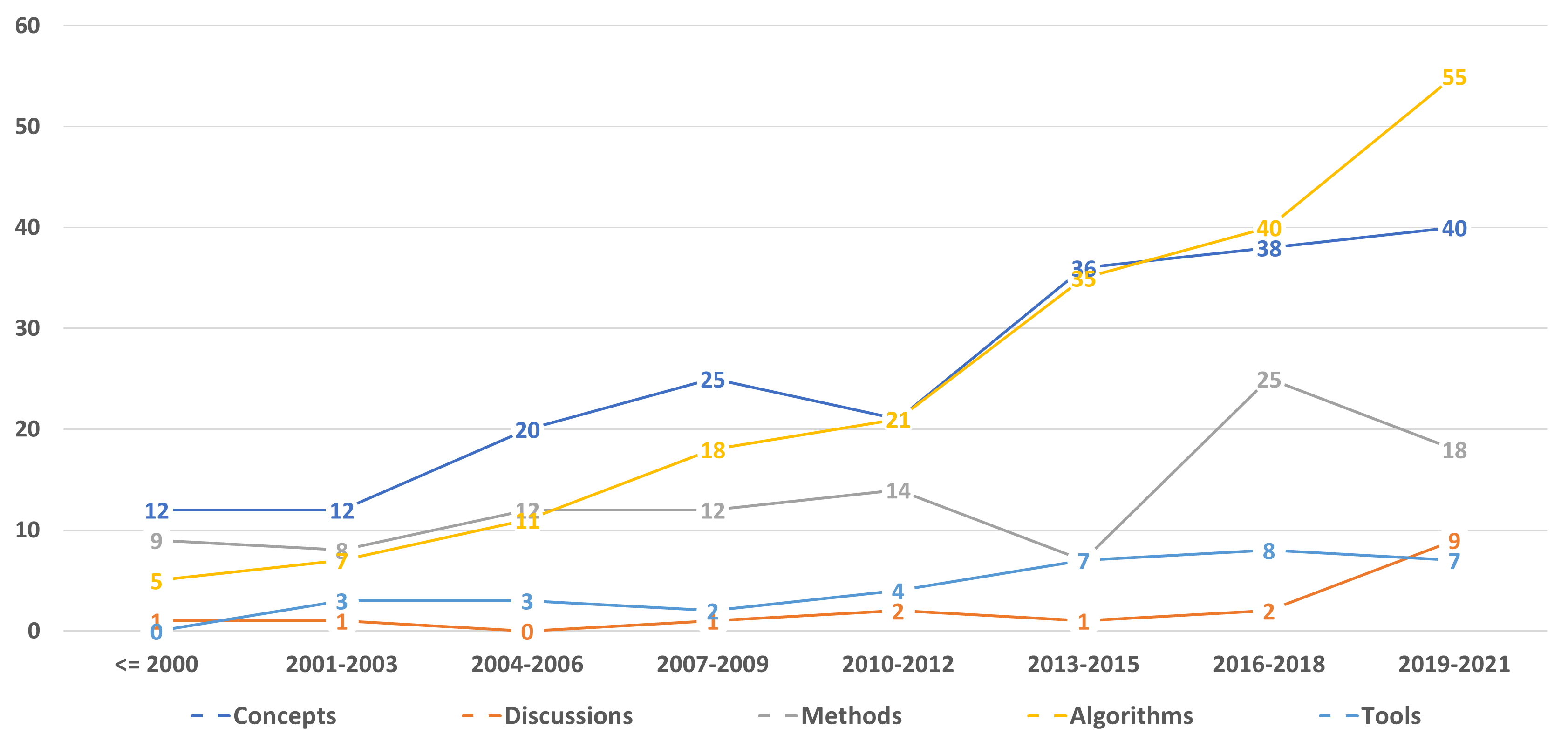}
    \caption{Number of publications classified based on contribution type and grouped by intervals of three years.}
    \label{fig:contribution-type-trend}
    \vspace{-.35cm}
\end{figure}

\paragraph{Contribution type.}
\label{sec:ContentAnalysis:Meta:ContributionType}
The main contribution types of CMAI research are concepts (204 occurrences), followed by algorithms (192 occurrences), methods (105 occurrences), tools (34 occurrences), and discussions (17 occurrences). Although tools are only presented as an explicit research contribution to a limited extent, we observed that 294 papers in the dataset use some form of tool support, of which 19.7\% is openly available via a public link. Fig.~\ref{fig:contribution-type-trend} further shows that algorithms have been the dominant contribution type since 2016, which can be explained by the rising application of AI techniques as Machine Learning, Genetic Algorithms, NLP, etc. 
20\% of the papers in the dataset make multiple contributions, with the following occurrences: concepts \& algorithms (49 occurrences), concepts \& methods (27 occurrences), methods \& algorithms (10 occurrences), concepts \& tools (2 occurrences), methods \& tools (2 occurrences), algorithms \& tools (1 occurrence), and concepts \& methods \& algorithms (1 occurrence). These combinations show that multiple artifacts may be needed to address a broad problem, which also corresponds to the DSR perspective.

\subsubsection{CM Analysis}
\label{sec:ContentAnalysis:ModelingTaxonomy}
In the following, an analysis is performed concerning the used modeling language and the modeling purpose taxonomy (see Table~\ref{tab:CM-taxonomy}).

\begin{figure}[hb!]
    \centering
    \vspace{-.1cm}
    \includegraphics[width=.95\linewidth]{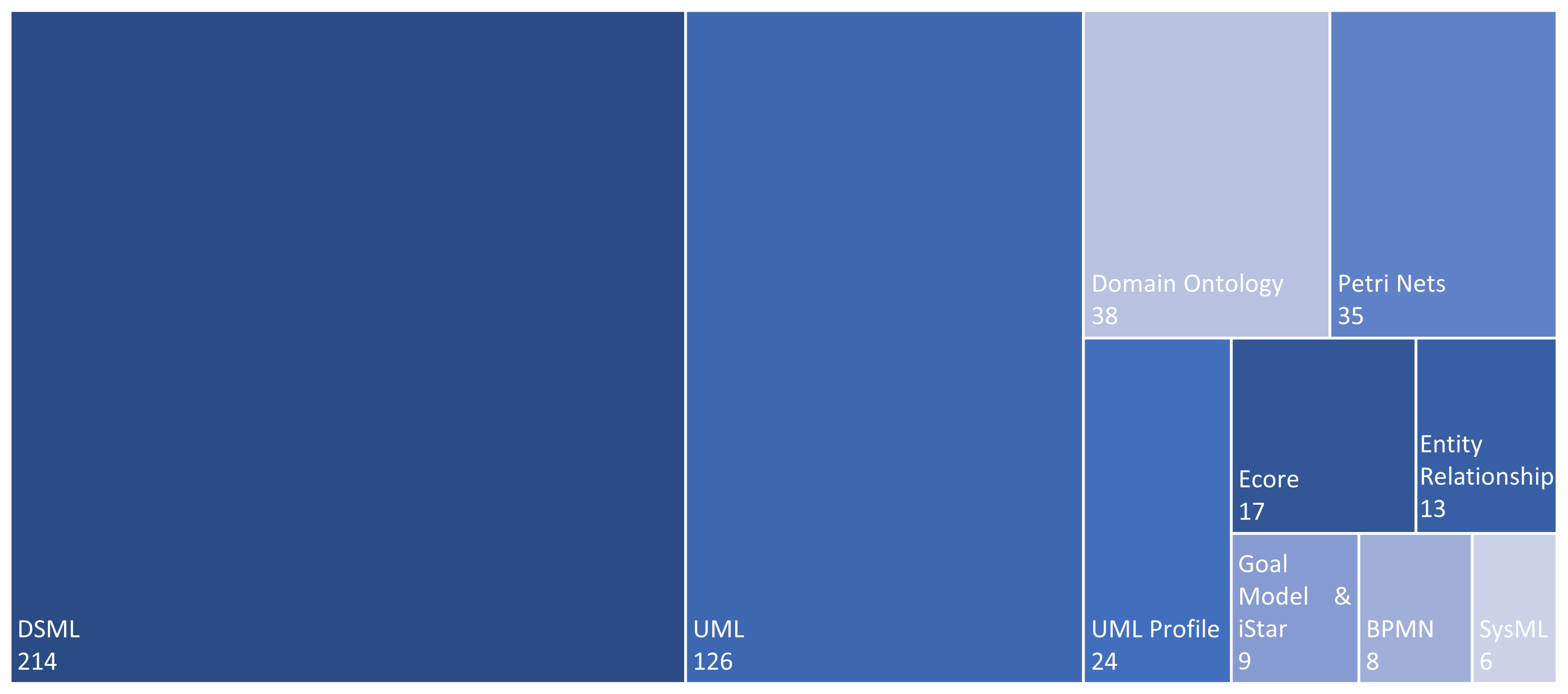}
    \caption{Publications share by modeling language (only languages with more than five occurrences are included).}
    \label{fig:modeling-language-treemap}
    \vspace{-.3cm}
\end{figure}

\paragraph{Modeling Language.}
The publication share per modeling language (group) is given in Fig.~\ref{fig:modeling-language-treemap}, which includes all languages with more than five occurrences in the dataset. These languages represent 90.6\% of the modeling languages and 87.8\% of the publications in the survey data. The modeling languages are roughly evenly distributed between being specific to the targeted problem domain (i.e., Domain-Specific Modeling Languages (DSMLs), Domain Ontologies, and UML (i.e., Unified Modeling Language) Profiles with 276 occurrences), and the following general-purpose modeling languages (i.e., a total of 214 occurrences): UML, Petri Nets, Ecore (i.e., part of the Eclipse Modeling Framework), Entity Relationship models, Goal models, Business Process Modeling Notation (BPMN), and SysML.

\begin{figure}[h!]
    \centering
    \includegraphics[width=.95\linewidth]{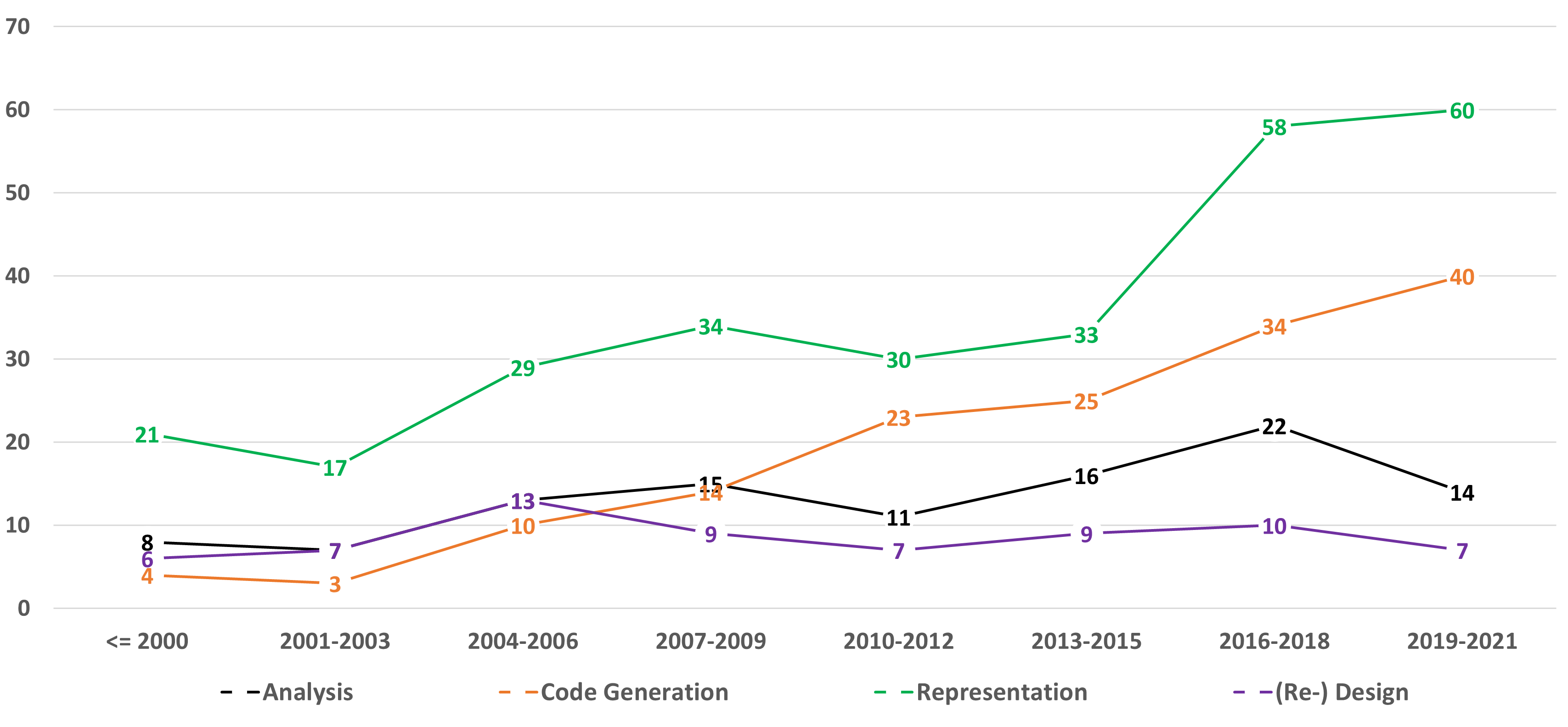}
    \caption{Number of publications classified based on modeling purpose and grouped by intervals of three years.}
    \label{fig:modeling-purpose-trend}
\end{figure}

\paragraph{Modeling Purpose.}
Fig.~\ref{fig:modeling-purpose-trend} gives an overview of the modeling purpose in the CMAI research papers. The most occurring purpose over time is the use of conceptual models to create a descriptive representation of the system under study (282 occurrences). Notably, the purpose of representation is often a precursor for the other purposes, as manifested by the following observed combinations: representation \& analysis (72 occurrences), representation \& (re-) design (26 occurrences), representation \& (re-) design \& analysis (41 occurrences), and representation \& code generation (6 occurrences). Since 2010, model-based code generation has gained importance in the CMAI field and currently occupies the second position with an incidence of 153. The use of CM for simulation or query analysis (106 occurrences) and (re-)designing a future version of the system (68 occurrences) remains relatively stable with only slight fluctuations over time.

\subsubsection{AI Analysis}
\label{sec:ContentAnalysis:AITaxonomy}
The analysis of the AI domain follows subsequently. First, the results for the mapping to the AI domain and subdomain are reported, followed by the mapping results concerning the value AI brings to CM.

\begin{figure}[h!]
    \centering
    \includegraphics[width=.95\linewidth]{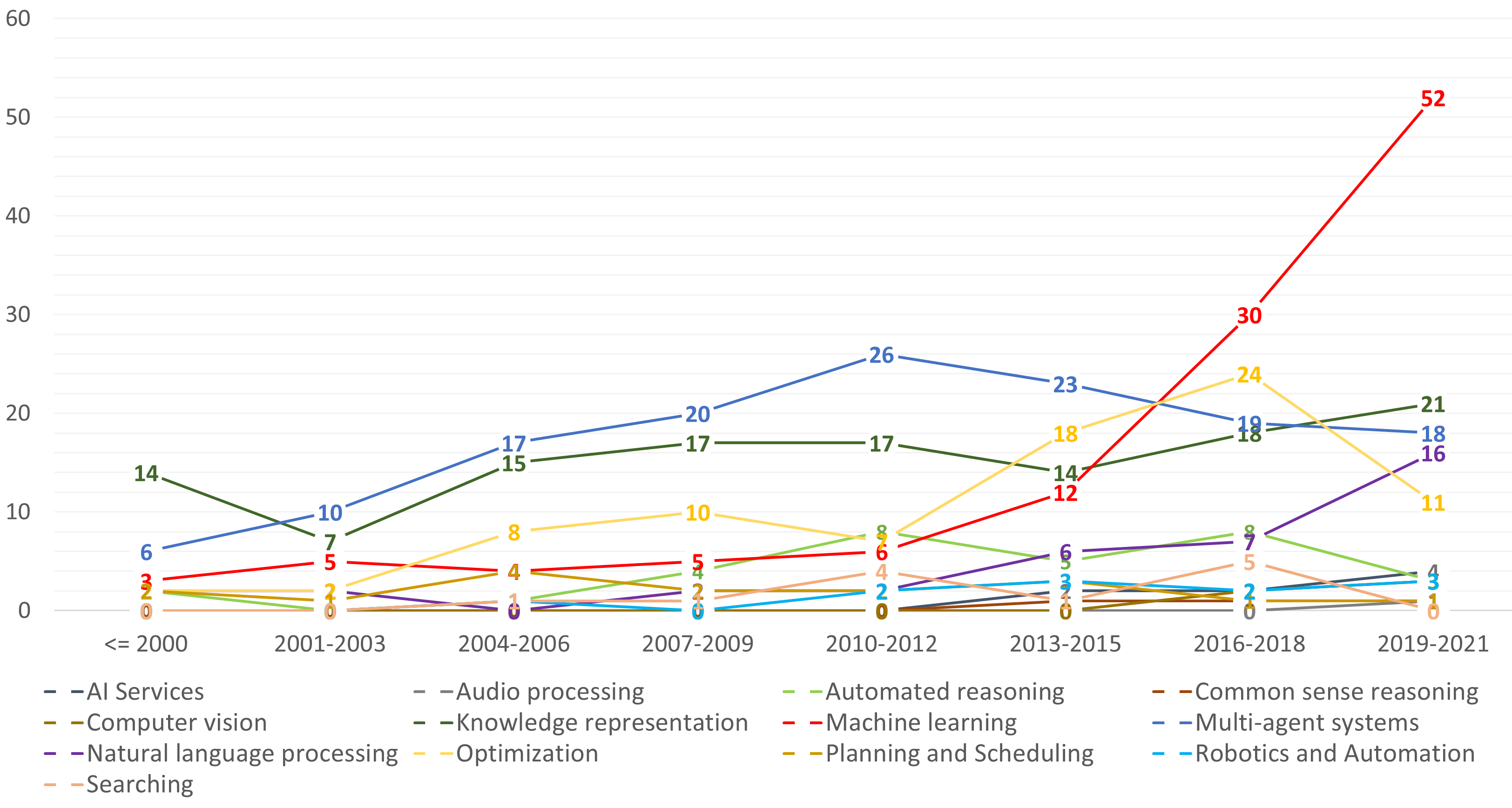}
    \caption{Number of publications classified based on the AI subdomain and grouped by intervals of three years.}
    \label{fig:AI-subdomain-trend}
\end{figure}

\paragraph{AI subdomain.}
Note that the AI subdomains Connected and Automated vehicles, AI Ethics, and Philosophy of AI were not found in the dataset and are therefore not represented in Fig.~\ref{fig:AI-subdomain-trend}. The occurrences of all remaining 13 (of the 16 introduced) AI subdomains (cf. Table~\ref{tab:AI-taxonomy}) are presented in Fig.~\ref{fig:AI-subdomain-trend}. Over the complete dataset, the AI subdomains can be classified according to the frequency with which they have been addressed in the CMAI domain: Multi-agent systems (139 times), Knowledge representation (123 times), Machine learning (117 times), Optimization (82 times), NLP (37 times), Automated reasoning (31 times), Planning and Scheduling (16 times), Searching (12 times), Robotics and Automation (11 times), AI Services (8 times), Computer vision (5 times), Common sense reasoning (3 times), Audio processing (1 time).
When analyzing the evolution, the fast growth of the Machine Learning domain since 2016 stands out. Besides, a slight increase can be observed for NLP use since 2019. Nevertheless, this rise in importance needs to be confirmed in the future. This is in contrast to the other domains, which show a flattening pattern over time -- albeit the increase in the total number of publications (cf. Fig.~\ref{fig:papers-per-year}).

\begin{figure}[h!]
    \centering
    \includegraphics[width=.95\linewidth]{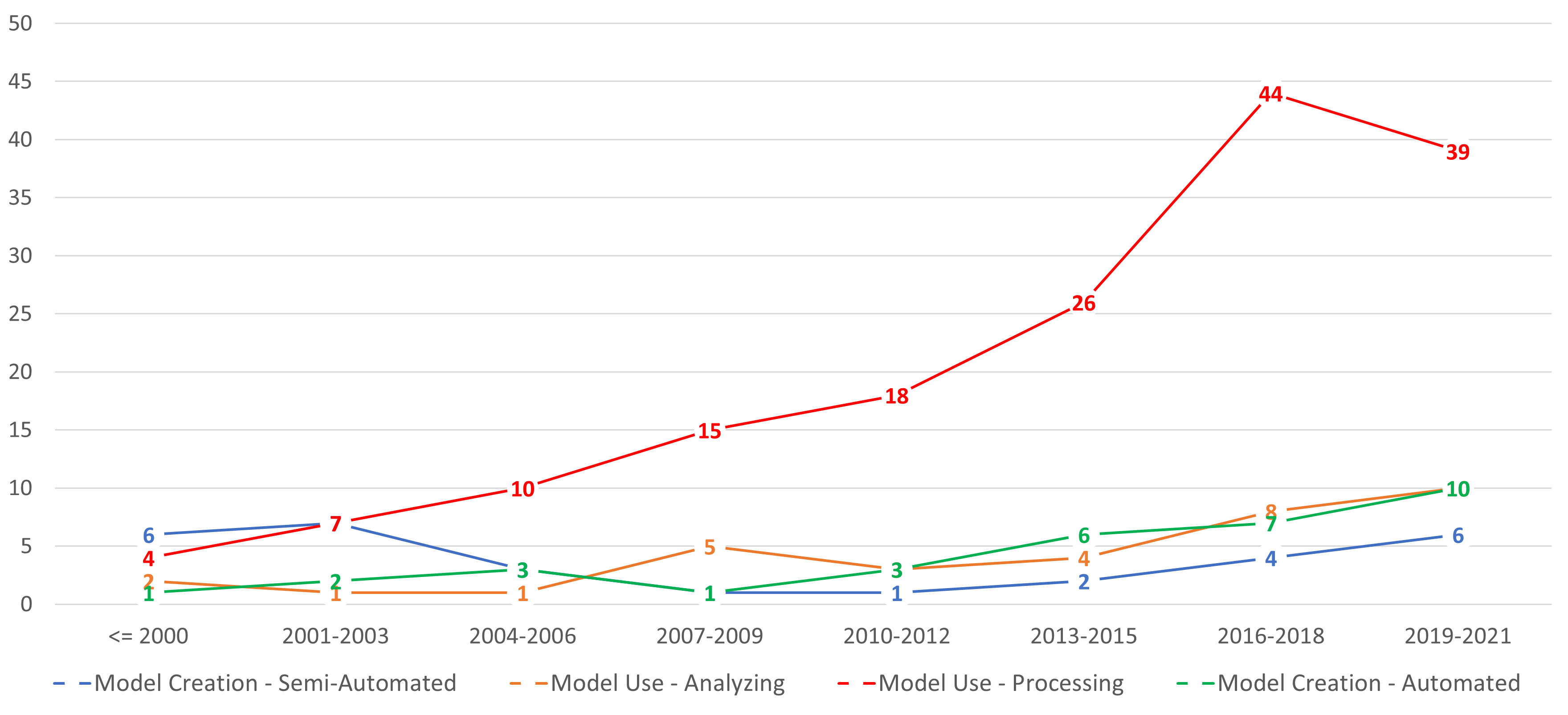}
    \caption{Number of publications classified based on the AI value to CM and grouped by intervals of three years.}
    \label{fig:AI-Value-To-CM-Trend}
\end{figure}

\paragraph{AI value to CM.}
Fig.~\ref{fig:AI-Value-To-CM-Trend} shows that AI brings value to the CM domain in 254 papers. The main benefit can be categorized as Model Use - Processing, which was observed 163 times. This category has been clearly dominant since 2004, followed by Model Use - Analysis (34 occurrences), Model Creation - Automated (33 occurrences), and Model Creation - Semi-Automated (30 occurrences). The occurrence of these benefits fluctuates between 1 and 10 over the 3-year intervals, so the mutual importance of these benefits changes regularly.

\subsubsection{CM and AI Analysis}
\label{sec:ContentAnalysis:CMAI}
In the following, we report on the findings on the combination of CM and AI. We first analyze the combination between modeling language and AI technique before combining three taxonomies to further dig into the most occurring combinations in CMAI research. 

\begin{figure}[h!]
    \centering
    \includegraphics[width=.99\linewidth]{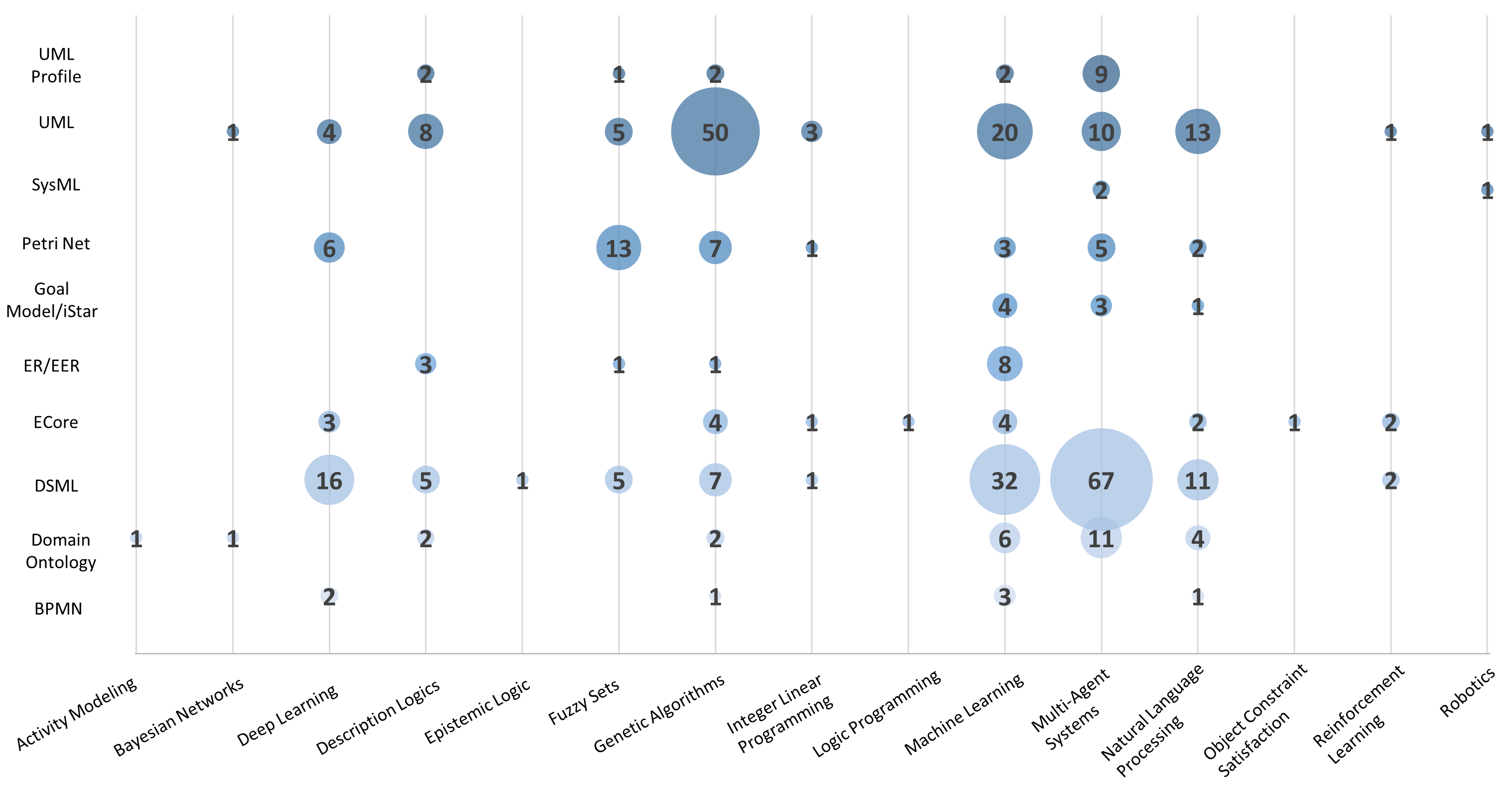}
    \caption{Use of Modeling Languages in combination with AI Techniques}
    \label{fig:Modeling-Language-Vs-AI-Technique}
\end{figure}

\paragraph{CM language and AI technique.}
Fig.~\ref{fig:Modeling-Language-Vs-AI-Technique} gives an overview of the combined use of modeling languages and AI techniques. The most popular combination is using DSMLs with Multi-Agent Systems (67 occurrences). This type of modeling language is also commonly used with Machine Learning (32 occurrences), Deep Learning (16 occurrences), and NLP (11 occurrences). Another language that stands out is UML, which is combined with AI techniques as Genetic Algorithms (50 times), Machine Learning (20 times), NLP (13 times), and Multi-Agent Systems (10 times). Other remarkable combinations are the use of Petri Nets with Fuzzy sets (13 occurrences) and the one of a Domain Ontology with Multi-Agent Systems (11 occurrences). All remaining combinations in Fig.~\ref{fig:Modeling-Language-Vs-AI-Technique} have less than ten occurrences in our dataset.

\begin{figure}[h!]
    \centering
    \includegraphics[width=.99\linewidth]{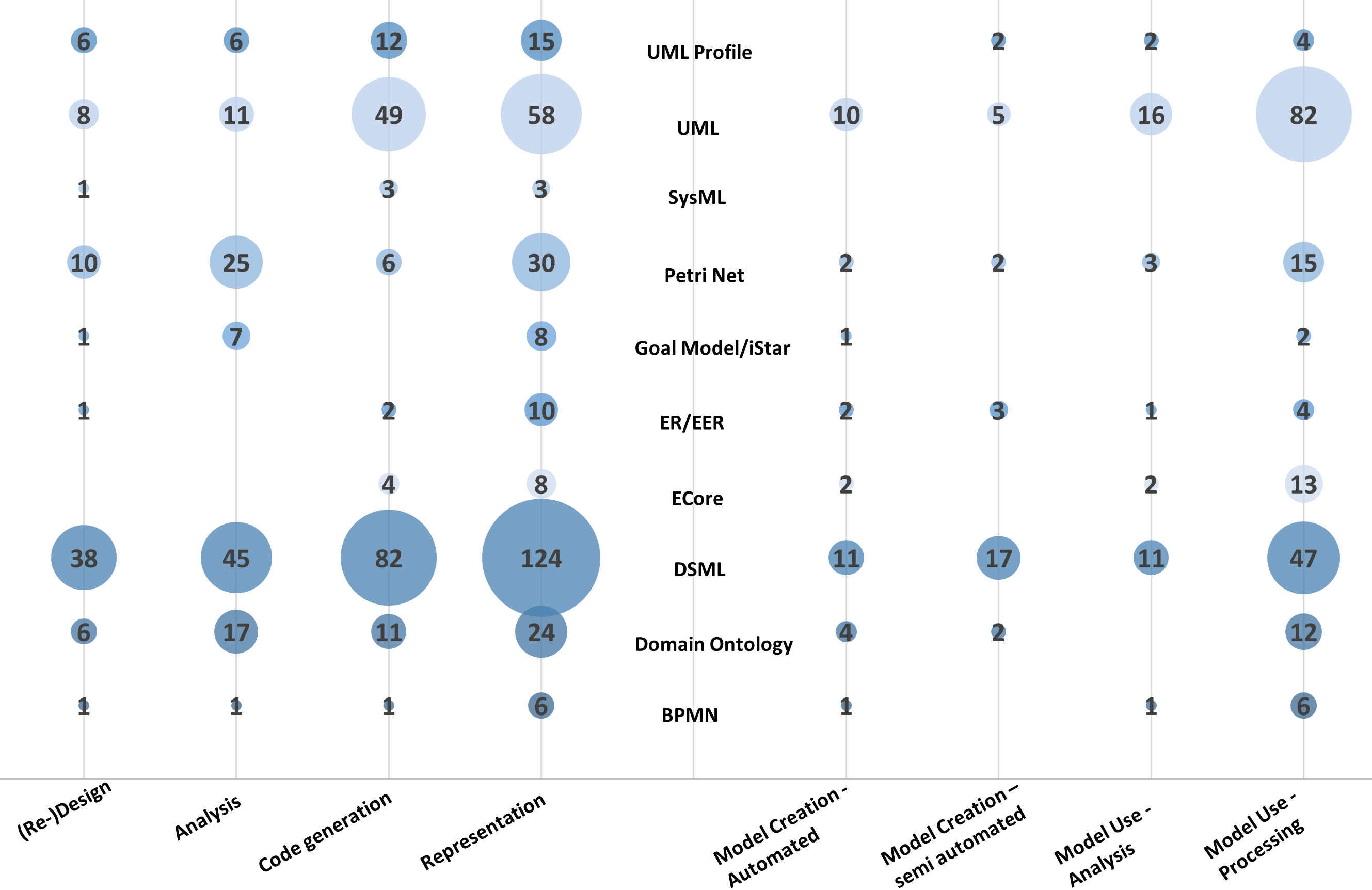}
    \caption{Overview of the modeling language combined with the modeling purpose and the AI value to CM.}
    \label{fig:Modeling-Language-Vs-CMAI}
\end{figure}

\paragraph{Modeling language, modeling purpose, and AI value to CM.}
Fig.~\ref{fig:Modeling-Language-Vs-CMAI} shows the most commonly used CM languages, their modeling purpose, and the value AI brings to CM. In line with the analysis in Fig.~\ref{fig:AI-Value-To-CM-Trend}, we see that Model Use - Processing is the primary value offered by AI to CM. In particular, this value is mainly realized in the context of UML (82 occurrences) and DSMLs (47 occurrences). Although the other benefits of AI to CM are less frequent, UML and DSMLs are also mainly employed to realize these. This goes hand in hand with the modeling purpose (see right side of Fig.~\ref{fig:Modeling-Language-Vs-CMAI}), which shows that DSMLs and UML are applied for the different modeling purposes with a total incidence of respectively 289 and 126. This demonstrates broader usability compared to the other languages in the dataset.

\begin{figure}[h!]
    \centering
    \includegraphics[width=.99\linewidth]{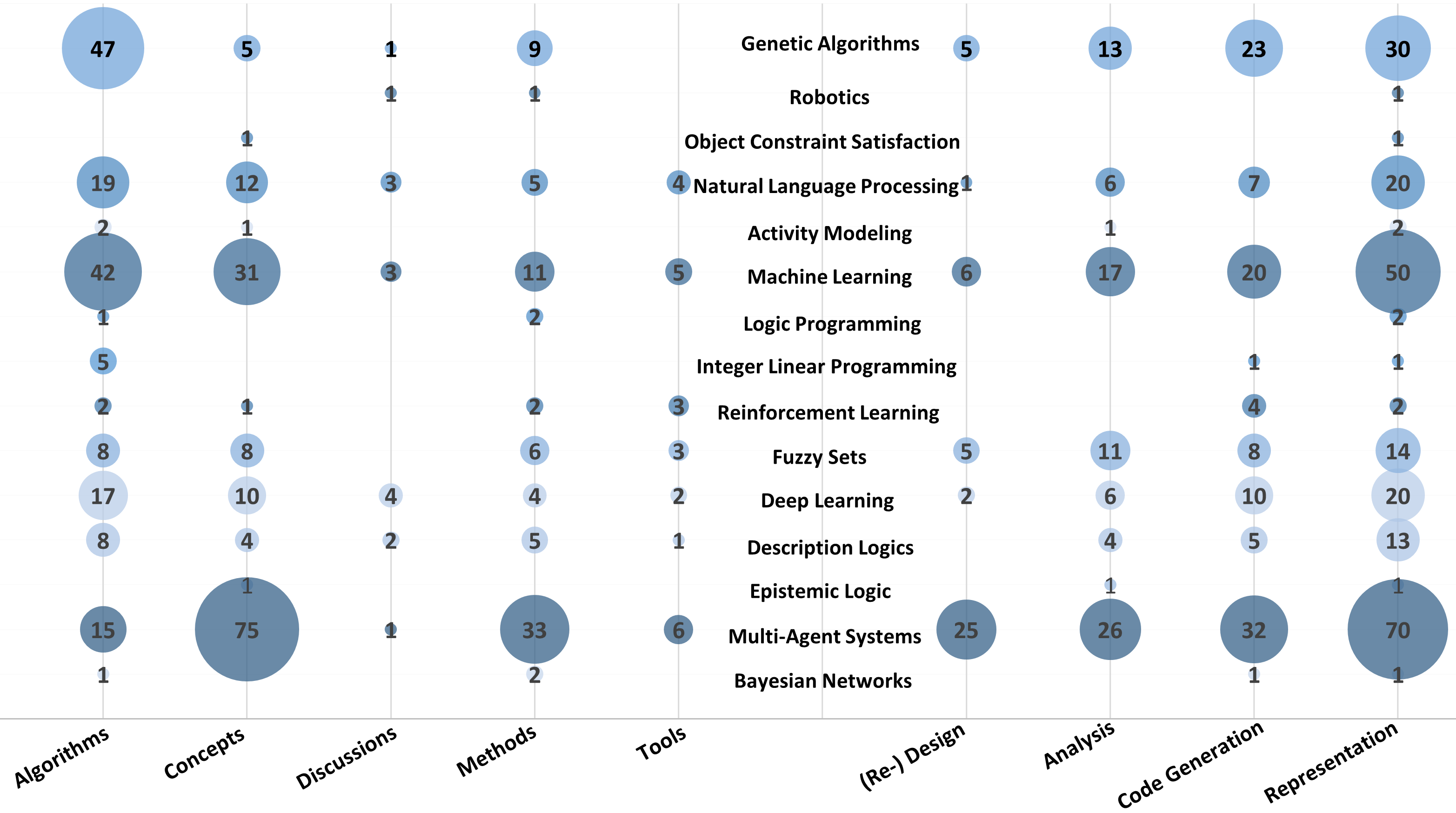}
    \caption{Overview of the AI techniques in combination with the contribution type and modeling purpose.}
    \label{fig:AI-ModelingPurpose-ContributionType}
\end{figure}

\paragraph{AI technique, contribution type, and modeling purpose.}
Fig.~\ref{fig:AI-ModelingPurpose-ContributionType} gives an overview of the contribution type and modeling purpose in combination with the different AI techniques. Overall, it can be observed that algorithms are developed as the main contribution for most AI techniques, such as Genetic Algorithms, Machine Learning, NLP, Deep Learning, etc. These algorithms are mainly combined by descriptive models representing the system under study.
Besides, concepts and methods are particularly relevant for model representation in the context of Multi-Agent Systems and Machine Learning. For these techniques, the languages also support the purposes of analysis, code generation, and (re-)design. 
The development of tool support as an explicit research contribution is still immature in the current CMAI research, so frequently occurring combinations cannot yet be identified.

\subsection{Research Community Analysis}
\label{sec:findings:Community}
\begin{figure}[ht]
    \vspace{-.1cm}
    \centering
    \includegraphics[width=.9\linewidth]{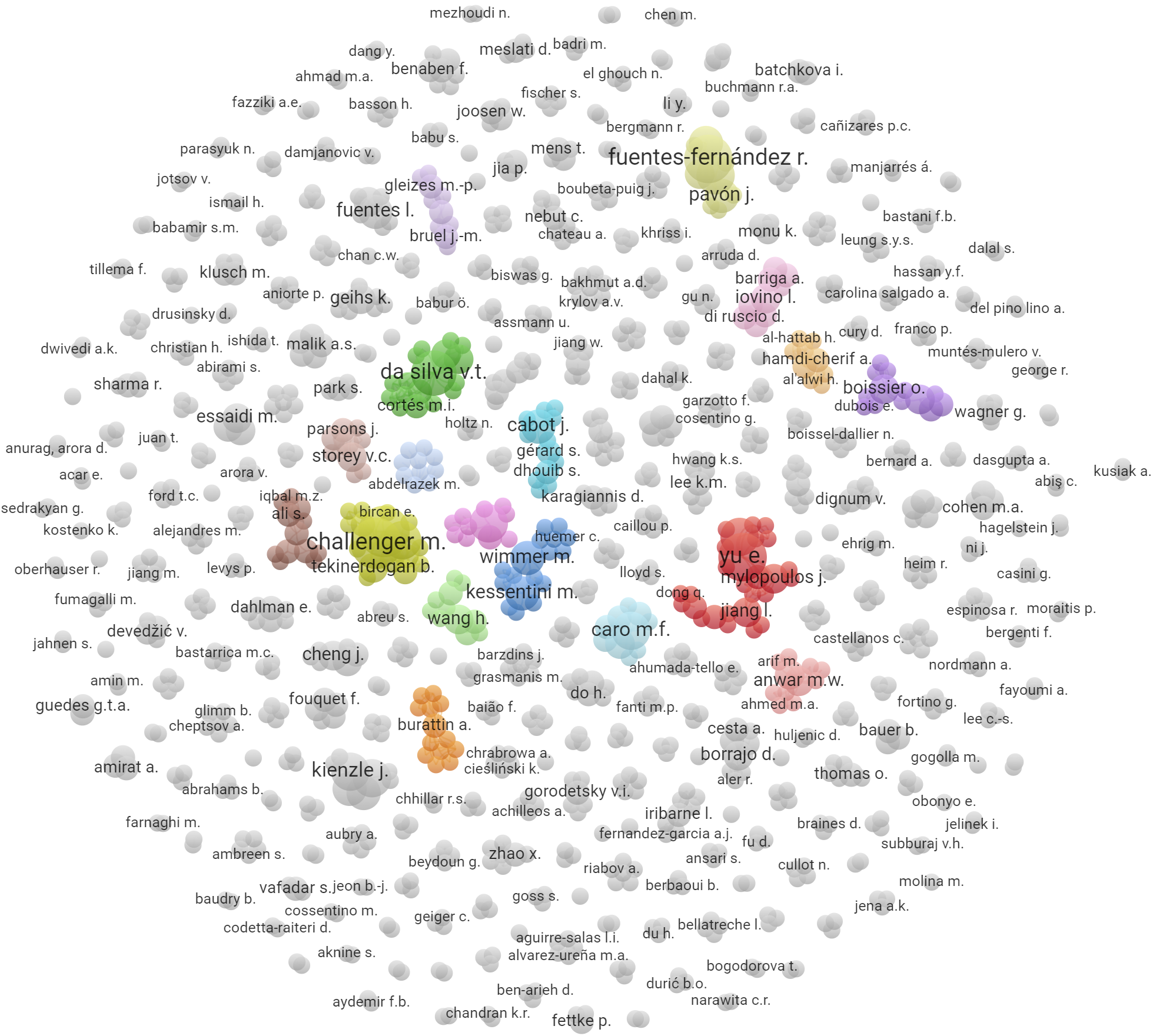}
    \caption{Co-authorship graph for the entire database of 460 publications (created with VOSviewer)}
    \label{fig:co-occurrence-graph}
    \vspace{-.3cm}
\end{figure}

In the following, we present the results of a co-authorship analysis performed on our dataset aimed at responding to RQ-6. Fig.~\ref{fig:co-occurrence-graph} shows the research communities contributing to the domain of CMAI. It shows all $1,186$ authors in our dataset and all co-authorships. Nodes in the graph represent authors, and edges between nodes represent co-authorship, i.e., connected nodes have at least one co-authored paper. In total, the analysis yielded 330 communities comprising between 1 and 28 members (i.e., collaborations through a co-authored paper in our dataset).

\begin{figure}[h!]
    \begin{subfigure}[b]{0.51\textwidth}
         \centering
         \includegraphics[width=\textwidth]{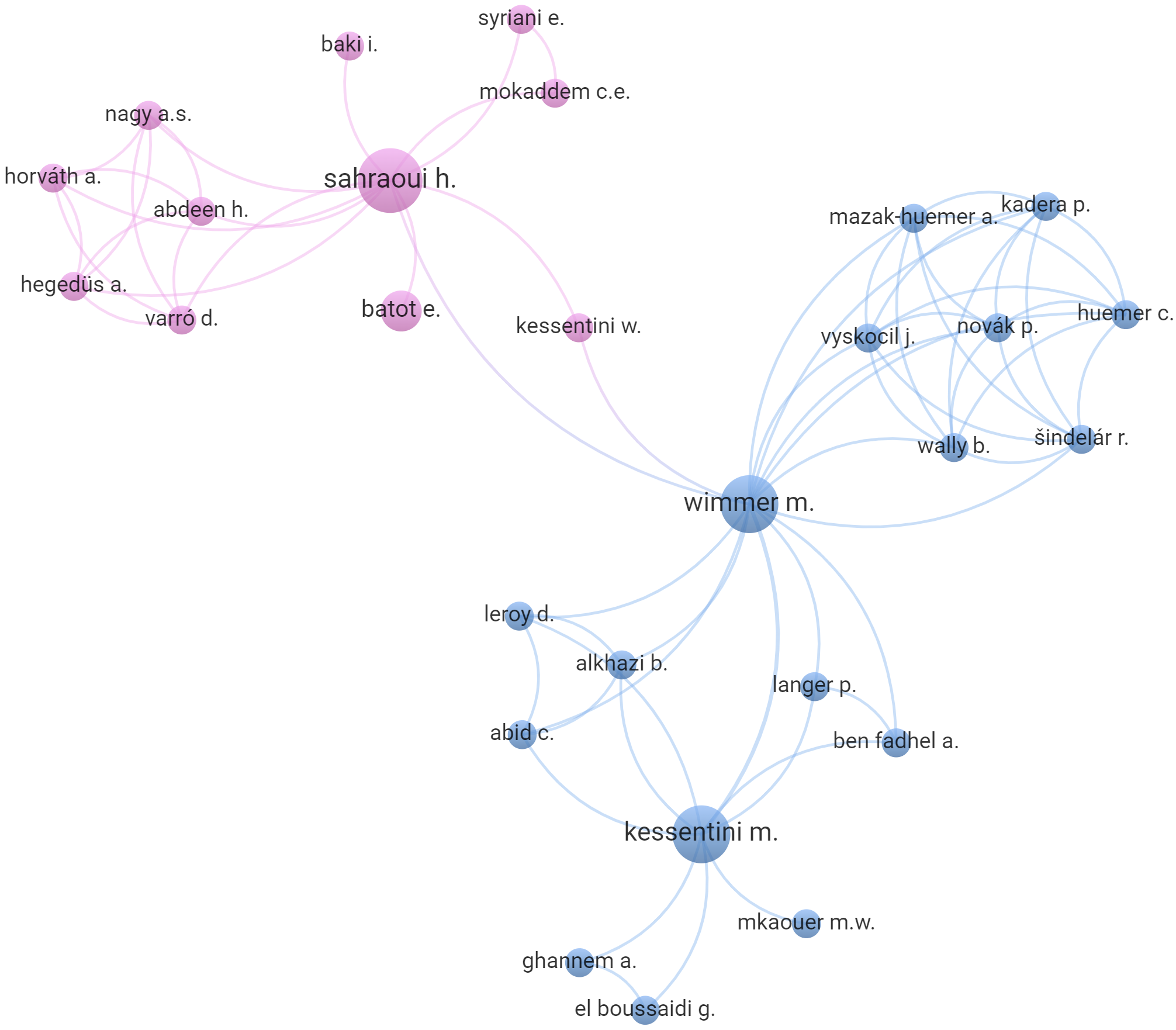}
         \caption{Manuel Wimmer, Houari Sahraoui community}
         \label{fig:occurrence-cluster-wimmer}
         \vspace{-.2cm}
     \end{subfigure}
     \hfill
     \begin{subfigure}[b]{0.48\textwidth}
         \centering
         \includegraphics[width=\textwidth]{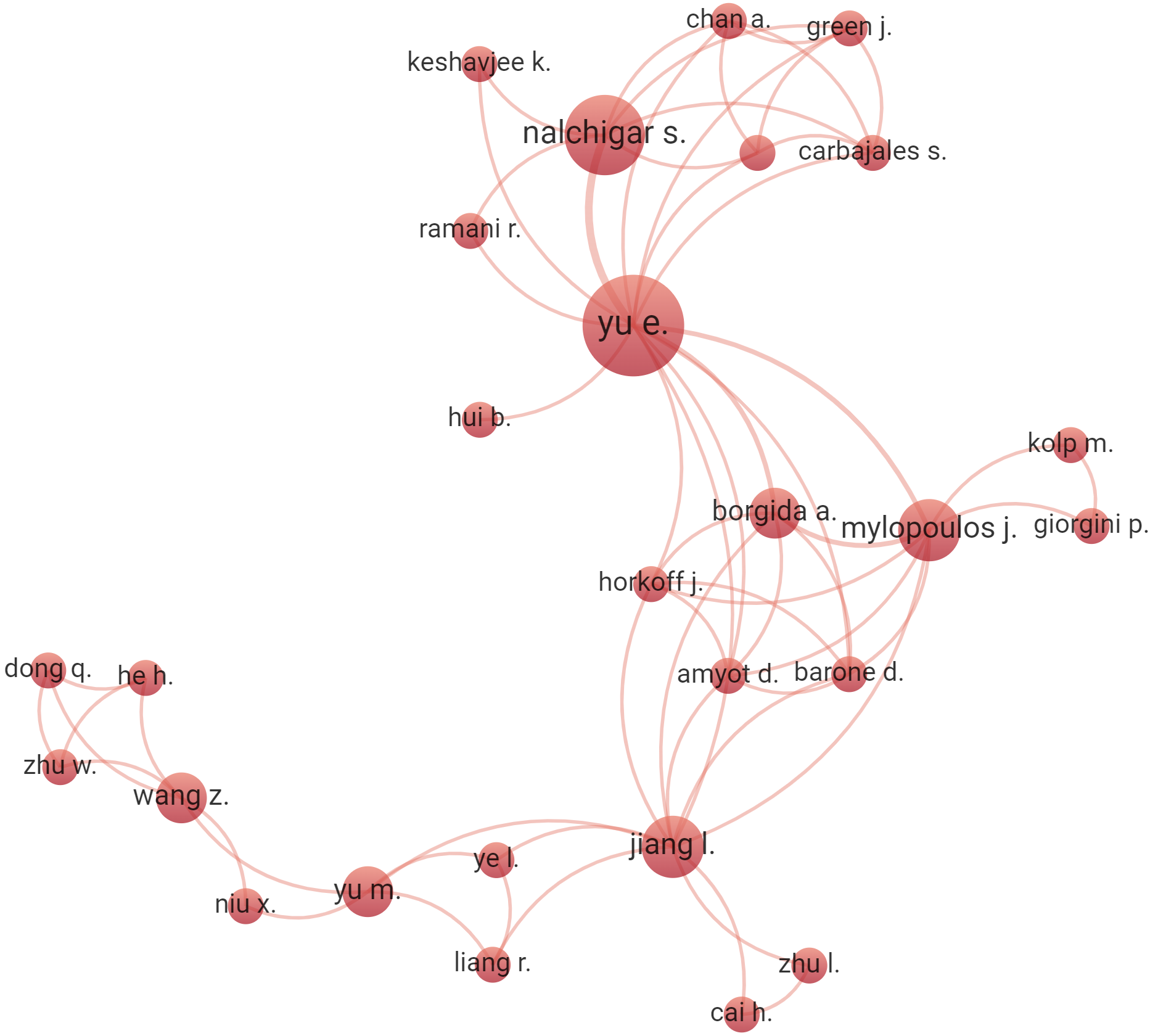}
         \caption{Eric Yu community}
         \label{fig:occurrence-cluster-yu}
         \vspace{-.2cm}
     \end{subfigure}
     \caption{The two most active CMAI research communities (representations created with VOSviewer)}
     \label{fig:three graphs}
     \vspace{-.3cm}
\end{figure}

When digging into the community analysis, we were interested in identifying the most active CMAI research communities in our dataset based on $(i)$ the size of the community (i.e., the number of nodes), and $(ii)$ the strength of the community (i.e., the number of co-authored documents). The sub-communities sharing more than two authors are present in the same color, while communities sharing a bridge author or bridge edge (like the community visualized in Fig.~\ref{fig:occurrence-cluster-wimmer}) have different colors. We ranked the communities first based on their size, followed by their strength. Table~\ref{tab:active-research-communities} summarizes the five most active communities. Each community is described by its central author(s), the total number of authors, and the total number of publications. We furthermore used the dataset and particularly analyzed the papers authored by these five communities. Consequently, Table~\ref{tab:active-research-communities} also shows the specific contributions and focus of the five most active CMAI research communities. In the following, we will briefly describe these five communities, Figs.~\ref{fig:occurrence-cluster-wimmer} and~\ref{fig:occurrence-cluster-yu} further visualize the two most active CMAI research communities.

\begin{table}[h!]
\centering
\caption{Biggest and most active research communities}
\label{tab:active-research-communities}
\footnotesize
\begin{tabularx}{\linewidth}{Xp{.1\linewidth}p{.124\linewidth}p{.18\linewidth}p{.23\linewidth}}
\toprule
\textbf{Community Central Authors} &\textbf{\# of Authors} &\textbf{\# of Publications} &\textbf{AI Domain} &\textbf{CM Domain} \\
    \midrule
     Houari Sahraoui, Manuel Wimmer  & 28 & 10 & Genetic Algorithms, & Representation \\
      & & & Machine Learning\\
     \midrule
     Eric Yu, Li Jiang, John Mylopoulos, & 27 & 12 & Machine Learning, NLP, & Representation \\
     Zhixue Wang& & & Description Logics\\
     \midrule
     Kardas Geylani, Moharram Challenger & 17 & 10 & Multi-Agent Systems, & Code Generation \\
     & & & Knowledge Representation\\
     \midrule
     Vivianne Torres Da Silva & 17 & 7 & Multi-Agent Systems, & Code Generation, Representation\\
     & & & Knowledge Representation \\
     \midrule
     Rubén Fuentes-Fernández & 13 & 8 & Multi-Agent Systems, & Code Generation\\
     & & & Knowledge Representation\\
    \bottomrule
\end{tabularx}
\end{table}

\begin{description}
    \item[Community centering Houari Sahraoui and Manuel Wimmer]: This community consists of 28 authors with ten documents in our dataset. The community has two sub-communities centered at Manuel Wimmer and Houari Sahraoui. Houari Sahraoui co-authored five publications, Manuel Wimmer is co-author of four publications in our dataset. These two communities work primarily on CMAI with Genetic Algorithms in the AI domain used for Code Generation and Representation in the modeling domain. This community primarily uses Ecore and UML Class Diagrams as modeling language. The analysis further showed that Wael Kassentini bridges the two sub-communities. 
    \item[Community centering Eric Yu, Li Jiang, John Mylopolous, and Zhixue Wang] This community consists of 27 authors who contributed 12 documents to our dataset. It comprises four sub-communities centered at Eric Yu, Li Jiang, John Mylopolous, and Zhixue Wang. Eric Yu is the most active author for this community, co-authoring eight publications. This community is primarily focused on Machine Learning for the AI domain and Representation \& Analysis for the CM domain. Some works are also related to Multi-Agent Systems and Automated Reasoning for the AI domain and Code Generation and Analysis for the CM domain. The most common combination is using the Goal Model language with Machine Learning. Goal Models and other DSMLs are used for representation analysis, while Machine Learning is primarily used to support Model Processing and Model Analysis. Mylopolous and Wang further focus on Automated Reasoning and on using DSMLs for Code Generation. 
    \item[Community centering Kardas Geylani and Moharram Challenger] This community consists of 17 authors with ten publications. Kardas Geylani and Moharram Challenger are the most active authors for this community, co-authoring eight publications. This community is primarily focused on Knowledge Representation with Multi-Agent Systems and using modeling for Code Generation. DSMLs are used in several cases.
    \item[Community centering Vivianne Torres Da Silva]: This community consists of 17 authors with seven publications. Vivianne Torres Da Silva is the most active author for this community co-authoring all seven publications. This community is primarily focused on Multi-Agent Systems for the AI domain and Code Generation and Representation for the conceptual modeling domain. Some works are also related to Multi-Agent systems and Automated Reasoning. This community's most commonly applied modeling languages are UML Profiles and DSMLs, which are used for Representation, Analysis, and Code Generation. 
    \item[Community centering Rubén Fuentes-Fernández] This communtiy consists of 13 authors with eight publications in our dataset. Rubén Fuentes-Fernández is the most active author with seven publications. This community is very similar to the community centering Kardas Geylani and Moharram Challenger because it also focuses on Multi-Agent Systems and the use of DSMLs for Code Generation and Representation.
\end{description}

\section{CMAI Web Knowledge Base}
\label{sec:WebSystemCMAI}
\begin{figure}[t]
     \centering
     \includegraphics[width=.99\textwidth]{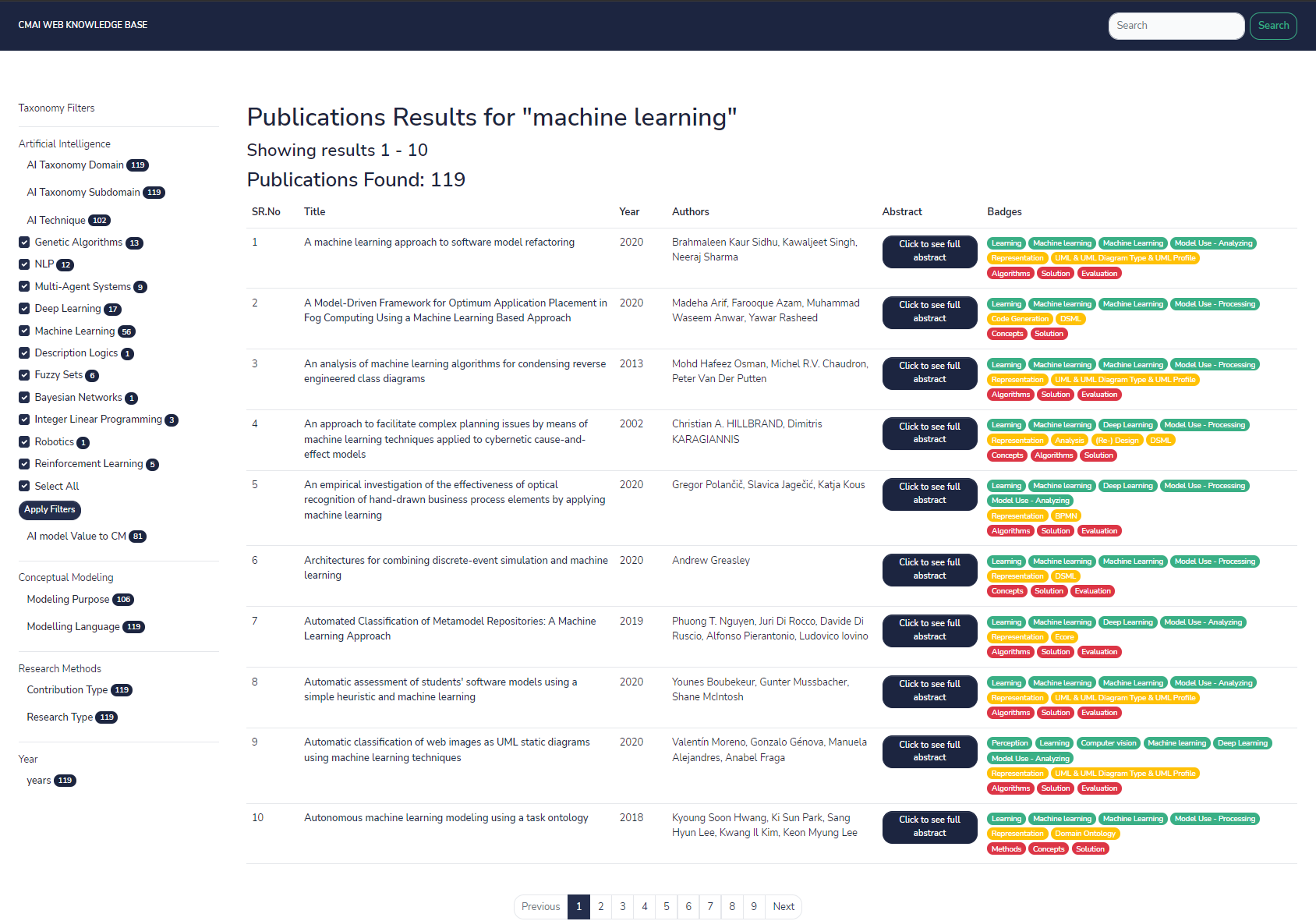}
     \caption{Publications results page of the CMAI web knowledge base}
     \label{fig:web-based-kb}
\end{figure}

In addition to the SMS we report with this article, we present a web-based knowledge base for our SMS that provides easy access to and enables efficient exploration of the publications, active authors, and prominent venues related to CMAI research. Our platform supports keyword-based search for publications, authors, and venues. All the searched publications are further tagged with the CM and AI taxonomies introduced in Sect.~\ref{sec:ResearchMethod:Phase4} (see Fig.~\ref{fig:web-based-kb}). The CMAI web knowledge base can be accessed via \url{http://me.big.tuwien.ac.at/cmai}\footnote{Please use the following credentials to log in, we will openly release the web knowledge base only with the publication of this article: U: j@g.com, PW: 123}.

The CMAI web knowledge base is intended to support the researchers interested in the CMAI domain and make the relevant resources easily accessible using our search platform. The platform initially supports the following use cases:
\begin{description}
    \item[Publications search] Our platform supports searching for publications based on keywords. The search query for publications is matched with the title and abstract of all publications in the CMAI SMS dataset. Fig.~\ref{fig:web-based-kb} shows the publications results page. The results page shows the total number of publications matching the search query. The title of each search result is a hyperref linking to the DOI of the paper, allowing users to navigate to the full text of a publication directly. The results page also allows the user to see the abstract of a search result by clicking the "see the abstract" button. 
    
    The user can also further filter the results using the three taxonomies: AI, CM, and Research Method (for research type and contribution type). The filtering allows multiple selections of taxonomy values, and the results are then evaluated based on the intersection of these taxonomy values. The user is also allowed to filter papers based on the publication year. By default, all the taxonomy values part of the search results are selected. In the left part of Fig.~\ref{fig:web-based-kb}, one can see the details of the AI technique taxonomy with all taxonomy chosen values and the corresponding number of results indicated. Each publication is further tagged with the corresponding taxonomy value (see the right side of Fig.~\ref{fig:web-based-kb}). 
    \item[Authors search] The CMAI web knowledge base allows users to search for the authors in our SMS by searching for author names in the query. The results page shows the total number of authors matching the search query. Furthermore, all publications in the dataset that those authors co-authored are presented. The platform also supports the filtering of results based on the taxonomies of the SMS, publication year, and the affiliation country of the author.
    \item[Venue search] The CMAI web knowledge base allows users to search for relevant venues matching a search query. The results show the total number of venues matching the query and all the publications in the dataset that have been published at this venue. The results can be further filtered using the taxonomies of the SMS.
\end{description}
Note that for all three uses cases, the platform also supports the search with an empty string, then showing all results of the dataset.

\section{Implications for Future Research}
\label{sec:TrendsAndVision}

Based on the insights of Sect.~\ref{sec:Findings}, we can derive the following implications for future research. First, the bibliometric analysis shows that CMAI research is topical, given the growing interest in the domain since 2015. This is also evinced by a panel discussion with the topic 'Artificial Intelligence meets Enterprise Modelling' at the Forum of the Practice of Enterprise Modeling (PoEM) workshop in 2019~\cite{Snoeck.19} and several workshops organized at international CM conferences (e.g., the CMAI workshop at Modellierung (since 2020)~\cite{Reimer.20}, the MDE Intelligence workshop at MoDELS (since 2019)~\cite{Burgueno.19-MDEIntelligence}, the CMAI workshop at ER (since 2020)~\cite{Bork.20CMAIWS}, and the International Workshop on Business Process Innovation with Artificial Intelligence (since 2017)~\cite{DeMasellis.17-BPAIws} at BPM. This indicates that current CMAI research is primarily CM-driven, in which the CM techniques are (re-)designed by existing AI techniques to target a specific problem. This observation shows that much potential is left for the CM community to provide value by (re-)designing AI techniques in combination with the use of CM (i.e., AI-driven research). In this respect, the CMAI web knowledge base (Sect.~\ref{sec:WebSystemCMAI}) supports researchers interested in the different aspects of the CMAI domain and makes the relevant resources easily accessible using the search functionality. 

The meta-analysis indicates that CMAI research is very suited to be guided by a Design Science Research methodological perspective ~\cite{gregor2013positioning, Peffers.07, Hevner2010}, as problems are addressed by the realization of different contributions (i.e., concepts, methods, algorithms, and tools) and iterative build-and-evaluate loops are needed. As such, the application of a design cycle offers the opportunity to realize a research contribution that combines both scientific rigor and practical relevance~\cite{Hevner2010}. The combined analysis shows that concepts and methods are particularly relevant for model representation in the context of Multi-Agent Systems and Machine Learning. At the same time, algorithms are developed as the main contribution for most AI techniques, such as Genetic Algorithms, Machine Learning, NLP, Deep Learning, etc. These algorithms are mainly combined by descriptive models representing the system under study.

Besides, Vision and Experience papers are highly welcomed to supplement current CMAI research efforts, as this type of research is currently still underrepresented in the CMAI domain. Given the current tool support, the meta-analysis also shows a need for openly available tools, which are presented as an explicit contribution of the research. This will foster the application of CMAI concepts, methods, and algorithms and reinforce the reproducibility of the presented results.

The content analysis of the papers reveals that DSMLs and UML are the mainly used modeling languages in the CMAI domain. Given their broader usability compared to the other languages, they provide an exciting starting point for future research in the field. Model-based code generation, Machine Learning, and NLP gained importance during the last decade, which offers an opportunity for future research in the CMAI domain. Besides, no research is currently available in the AI subdomains Connected and Automated vehicles, AI Ethics, and Philosophy of AI. It would be helpful to investigate whether multidisciplinary CMAI research projects could be beneficial to fill this gap.

\section{Threats to Validity}
\label{sec:Validity}
We will report on the threats to validity related to our SMS in the following. We will refer to the widely adopted generic set of validity threats proposed by Wohlin et al.~\cite{Wohlin.12}: \emph{Construct validity}, \emph{Internal validity}, \emph{External validity}, and \emph{Conclusion validity}. As we report on an SMS, the presented results reflect the investigated domain (i.e., CM and AI in our case) but naturally cannot be generalized. Consequently, external validity is not an issue, and the following discussion will concentrate on the remaining threats. Moreover, in the conclusion validity, we will relate to the specific reliability threats of mapping studies in Software Engineering that have been proposed in~\cite{Wohlin.13}.

Although SMSs are intended to provide an unbiased and complete overview of a specific domain of interest, such studies are still prone to validity threats. Most importantly, it needs to be stated that all results reported in this paper are only based on the sample of documents that $(i)$ were found by our search query and that $(ii)$ fit our inclusion and exclusion criteria. As soon as one of these parameters is altered, the results will also significantly change. As we aimed to provide a comprehensive overview of the CMAI domain, we designed the search query as inclusive as possible to enable a subsequent manual analysis and mapping. Consequently, we included search terms like 'intelligent' and 'smart,' which increased the initial hits with many papers that did not meet the inclusion/exclusion criteria and were dropped throughout the process. Still, this general query ensured we did not miss relevant work.

Looking at the inclusion and exclusion criteria, we focused on the intersection of CM and AI and high-quality, peer-reviewed works. In response to the latter argument, we omitted Google Scholar from the database selection and defined appropriate exclusion criteria (e.g., excluding non-scientific articles, imposing a minimum paper length, etc.). 

With respect to internal validity, it needs to be stated that our SMS only covers the published works, i.e., it is prone to the \emph{publication bias}. Moreover, we also only consider works written in English. Reliability also threatens the results presented in this paper. A study is reliable if the results are the same if the mapping is performed a second time. Here, we see the issue that it might not always be decidable, which taxonomy values can be mapped to a specific paper. Although we related our taxonomies to related works and established definitions, we arguable see a threat. Taxonomies like modeling purpose or AI modeling value have a subjective part, while many others like the modeling language and AI technique taxonomies are objective (and therefore considered more reliable). To mitigate this reliability threat, we conducted pilot mappings and conducted regular meetings throughout the mapping phase where we discussed the indistinct cases. Moreover, we set up a shared Google Docs spreadsheet with the pre-configured taxonomies. This ensured that all authors were able to track the entire mapping process, and it made sure that $(i)$ only valid taxonomy classes were mapped to a paper, and $(ii)$ that within one category only the proper taxonomy was applied.

To ensure replicability, we transparently followed a rigorous research methodology. We further developed and openly shared a web knowledge base for our SMS to trace the mapping process and reproduce the results we presented. We also mitigated the \emph{fishing problem}~\cite[p.\ 67]{Wohlin.12} because we were entirely open and curious about the state of research in the CMAI domain. Not a single paper of one of the authors is part of the dataset. Finally, all conclusions were drawn after the analysis, when the findings materialized.

\section{Conclusion}
\label{sec:Conclusion}
To the best of our knowledge, this article is the first to systematically and comprehensively analyze the state of research that combines CM and AI. Starting from $27,255$ documents, $460$ relevant ones have been identified, systematically analyzed, and mapped to respond to several research questions. With this SMS, we reported on the current state of CMAI research and the most used modeling languages and AI techniques in this domain. We further analyzed which combinations of modeling languages and AI techniques are most frequent and investigated the purpose and value of such a combination. The study at hand also showed the most active research communities and the publication channels that are used to publish CMAI research by applying network analysis. 

Our mapping study revealed a dominance of DSMLs and UML as the primary modeling languages used in CMAI research. Given their purpose- and stakeholder-specific design, DSMLs seem to provide an exciting starting point for many CMAI researchers. Especially, DSMLs are frequently combined with Multi-Agent Systems, Machine Learning, Deep Learning, while the UML is most often combined with Genetic Algorithms, Machine Learning, and NLP. We further found that quite some research exists in combining Petri Nets with AI; most often, these works combine it with Fuzzy Sets. AI is generally most often used to process the information provided by a conceptual model, whereas CM is most often used for representation with a recent update of code generation purposes.

Although the first works in CMAI date back to the late 1980s, the field is gaining more and more traction in recent years, most probably also influenced by the popularity of AI research. Our dataset shows that 48.9\% of the publications have been published since 2015. Since 2011, 67\% of all CMAI papers in our dataset have been published. With the expected continued increase of interest in CMAI research, we believe this article plays an essential role in structuring the state of research, showing vital research areas and areas that are still open for initial contributions. Developing the CMAI web knowledge base further enables efficient exploration of the CMAI domain. The platform is available via \url{http://me.big.tuwien.ac.at/cmai}.

\bibliographystyle{ACM-Reference-Format}
\bibliography{sample-base}

\end{document}